\documentclass[aps,pra,twocolumn,final,floatfix,superscriptaddress,10pt]{revtex4-1}
\usepackage[utf8]{inputenc}
\usepackage{physics}
\usepackage{mathtools}
\usepackage{amsmath}
\usepackage{xcolor}
\usepackage{tikz}
\usepackage{qcircuit}
\usepackage{soul}

\usetikzlibrary{graphs,graphs.standard}

\begin{document}

\title{Problem-Size Independent Angles for a Grover-Driven Quantum Approximate Optimization Algorithm}
\author{David Headley} \email{DavidKeithHeadley@gmail.com} \affiliation{Mercedes-Benz AG, Stuttgart, Germany}  \affiliation{Theoretical Physics, Saarland University, 66123 Saarbr{\"u}cken, Germany}

\author{Frank K. Wilhelm} \affiliation{Theoretical Physics, Saarland University, 66123 Saarbr{\"u}cken, Germany}\affiliation{Institute for Quantum Computing Analytics (PGI 12), Forschungszentrum J\"ulich, 52425 J\"ulich, Germany}
\date{\today}

\begin{abstract}
The Quantum Approximate Optimization Algorithm (QAOA) requires that circuit parameters are determined that allow one to sample from high-quality solutions to combinatorial optimization problems. Such parameters can be obtained using either costly outer-loop optimization procedures and repeated calls to a quantum computer or, alternatively, via analytical means. In this work we demonstrate that if one knows the probability density function describing how the objective function of a problem is distributed, that the calculation of the expectation of such a problem Hamiltonian under a Grover-driven, QAOA-prepared state can be performed independently of system size. Such calculations can help deliver insights into the performance of and predictability of angles in QAOA in the limit of large problem sizes, in particular, for the number partitioning problem.
\end{abstract}

\maketitle

\section{Introduction}

Variational Quantum algorithms aim to exploit the power of Noisy, Intermediate-Scale Quantum (NISQ)  computers \cite{preskill2018quantum} through the use of parameterised quantum circuits. The Quantum Approximate optimization Algorithm (QAOA) \cite{farhi2014quantum} provides a universal \cite{lloyd2018quantum, morales2020universality} ansatz state, efficiently preparable on a quantum computer, measurements from which might provide a useful heuristic method for the approximate solution of problems in combinatorial optimization.

QAOA circuits utilizing standard single-qubit-$X$ drivers recover the asymptotic scaling of Grover's algorithm \cite{GroverQAOANASA}. Furthermore, it has been shown in the same Grover-oracle problem context, that parameters in QAOA concentrate \cite{akshay2021parameter}. That is, for increasing problem size, the variance in the optimal angles for QAOA circuits tend to concentrate at terminal values \cite{rabinovich2021progress}. When considering QAOA in its original incarnation of single-qubit-$X$ drivers and problem Hamiltonians encoding combinatorial optimization problems such as Max-Cut, Max-$3$-Lin-$2$, Max-$k$-SAT, numerical studies have provided evidence that parameters also tend to concentrate, and, as such, can be determined for classes of similar problems without access to a quantum processing unit using classically tractable tensor network approaches \cite{streif2020training}. 

That optimal angles in QAOA concentrate, for certain classes of problem, is useful due to the exponential cost of using classical optimizers in the number of parameters over which they operate. Should QAOA provide a good heuristic method for a number of layers greater than $\approx 10$, it is likely that naive global classical optimization with repeated calls to the quantum computer to estimate and extremize the expectation value of the problem Hamiltonian is not the method used, due to the excessively large number of calls to the quantum computer required. Alternative methods to calculate optimal angles might use numerical methods on smaller or simplified problem sizes, as aforementioned, or use the averaged behavior of problems at limiting size. Using this second approach, Farhi et. al. demonstrated that for Sherrington Kirkpatrick type spin glasses and similar models, using single-qubit-$X$ drivers, one can average over model instances to find optimal angles in a calculation that is independent of the number of spins in a model \cite{farhi2019quantum}. This provides a formula that has been numerically optimized to a depth of $p = 10$, a result later generalized \cite{basso2021quantum} to show that the same technique extends to problems involving clauses incorporating more than two variables, providing an expression numerically tractable up to a depth $p = 20$ with the specific hardware used. This work was later widened to the context of Max-Cut problems on varying random graph families \cite{boulebnane2021predicting}, with Claes et. al. considering QAOA on infinite size mixed-spin SK models \cite{claes2021instance} and showing that the expected performance of QAOA also concentrates in this context at depth $1$.  

This paper takes an alternative approach to finding angles in the large problem-size limit. By using a Grover-style driver that bestows maximal permutation symmetry to the expectation value of a problem Hamiltonian under the QAOA state produced, we simplify the calculation of optimal angles at large $n$. This procedure requires that the probability density function (pdf), or density of states of the problem to be solved is known. In particular, we provide a method to find optimal angles and their corresponding expectation values on a classical computer, such that high quality solutions can be sampled from a quantum computer without the need for outer-loop optimization of the on-device QAOA state. We find that this method can be applied to the Number Partitioning Problem (NPP) and other problems where the probability density function is known or can be approximated.

The Ans\"atze used in variational quantum algorithms, in general, must have problem specificity to avoid the fate of barren plateaus \cite{mcclean2018barren, cerezo2021cost}. In QAOA, this problem specificity enters into the ansatz twice. Firstly, the problem Hamiltonian takes its eigenvalues directly from the objective function of the problem to be solved. Secondly, and more subtly, the driver Hamiltonian biases certain transitions between computational basis states over others. The use of Grover drivers removes the second of these and allows one to quantify where the performance in QAOA lies, between that of algorithms that use amplitude amplification-like structure agnostic speedups to solve problems, to those which derive performance from classically exploitable problem structure with respect to the single-qubit basis. 

Other works have used Grover drivers in the context of QAOA. It has been shown that combinatorial optimization problems can be approximately solved using Grover's algorithm via the compilation of their objective function to a threshold function. Grover's algorithm is asymptotically optimal for the Grover problem \cite{zalka1999grover}, with the performance of the threshold-based algorithm numerically observed to outperform the non-threshold version, however, incurring additional compilation overhead \cite{bartschi2020grover, golden2021threshold, golden2022evidence}. Though one might not a priori know which threshold to use for such a compiled version, knowledge of the pdf of the problem, as used in this work would certainly be sufficient knowledge for the selection thereof.

While it is not expected that Grover driven QAOA should provide superior performance to that of well suited problem specific drivers, it is expected that Grover driven QAOA will not suffer from the issue of exponentially diminishing gradients, barren plateaus---the angles can be obtained independently of problem size on a classical computer, provided the problem's pdf is known---in the case that the drivers are not well suited to the problem. As such, Grover driven QAOA provides a point of reference to the performance of a QAOA-type algorithm in the very average case of driver-problem synergy.

\label{section-DASCHEME}


\section{Theory}
QAOA is performed via $p$ applications of driver and problem Hamiltonians applied for parameterised times $\vec\beta, \vec\gamma$. Applying such Hamiltonians to a state prepared in an equal superposition of computational basis states, one produces:
\begin{equation}
\ket{\vec \beta,\vec\gamma}  =\prod_{p'=1}^p e^{i\beta_{p'}\hat H_{\rm D}} e^{i\gamma_{p'} \hat H_{\rm P}}\ket+^{\otimes n}, \label{QAOA_state}
\end{equation}
which is the problem dependent QAOA ansatz state, depending on $2p$ free parameters. One is then required to find angles such that when sampling from the computational basis measurement outcome distribution of the ansatz state, one finds configurations corresponding to small or large eigenvalues of $H_P$ depending on whether a problem is one of maximization or minimization. To this end, the expectation:
\begin{equation}
 E_{p}(\vec \beta,\vec\gamma) =\langle \hat H_{\textrm{P}} \rangle _{\vec \beta,\vec\gamma} = \bra{\vec \beta,\vec\gamma} \hat H_{\textrm{P}} \ket{\vec \beta,\vec\gamma},
\end{equation} 
should be extremized with respect to the variational parameters. Measurements of the optimised ansatz state then correspond to high quality solutions to a binary-variabled combinatorial optimization problem specifying $\hat H_P$. A combinatorial optimization problem on $n$ binary variables has objective function $C(z)$ with $z\in \{0,2 ^n -1\}$ from which the problem Hamiltonian is defined as
\begin{equation}
    \hat H_{\textrm{P}} = \sum_{z} C(z)\ket z\bra z
\end{equation}
but more typically written in a Hadamard-transformed clause representation as 
\begin{equation}
    \hat H_{\textrm{P}} = \sum_{k=0}^{2^n-1} g(k) Z^{k_1}Z^{k_2}... Z^{k_n}
\end{equation}
where $k_i$ is the $i$th bit of the binary representation of $k$. Typical problems for QAOA utilize low weight interactions such that compilation to near term quantum computers allows for state preparation within the coherence time of a device. Such problems include Erd\H{o}s R\'enyi random Max-Cut problems \cite{crooks2018performance} in which one takes 
\begin{equation}
    g(k) \sim  \begin{cases} \mathrm{Bernouli}(f) \,\, \text{if} \,\, \mathrm{weight}(k) = 2;\\
    0,\,\, \mathrm{otherwise}.
    \end{cases}    
\end{equation}
 for filling factor $f$ and $\text{weight(k)}$ denoting the bit-weight of the label $k$. SK models are defined similarly with weights being sampled from a Gaussian distribution rather than the Bernoulli distribution of Max-Cut \cite{mezard2009information}. Also of interest are max-$k$-SAT type problems \cite{akshay2020reachability} in which terms up to weight $k$ are present, taking integer values. Number partitioning problems, considered in section \ref{npp_section} consist of all-to-all coupled Mattis type Ising spin glasses in which the weights of $ZZ$ interactions take the product of random variables associated with the two constituent bits. The descriptions $g(k)$ and $C(z)$ are related via an $n$-dimensional binary-variable Fourier transform, otherwise known as the Walsh-Hadamard transform \cite{hadfield2018quantum}.
 

\subsection{Properties of Grover Drivers}

The driver Hamiltonian $\hat H_D$ typically consists of single-qubit $X$ operations,
\begin{equation}
\hat H_{D} = \sum_{i=1}^nX_i.
\end{equation}
Various works have investigated the use of drivers that differ from this standard single-qubit driver. For example, a version of QAOA which samples from a differing selection of higher order drivers for each layer has been shown to improve QAOA convergence \cite{zhu2020adaptive, chen2022much}. Drivers can be used to shift complexity from problem to driver Hamiltonian in the case of problems of hard and soft constraints \cite{hadfield2017quantum}.

In this work, an alternate driver---that which generates the Grover mixing operator---is used. The Grover driver can be written as a projector onto a Hadamard basis product state $\ket + ^ {\otimes n}$:
\begin{equation}
    \hat H_G =\ket+^{\otimes n}\bra+^{\otimes n} = \left[\frac{1}{\sqrt{2^n}} \sum_{z= 0}^{2^n-1} \ket z \right]\left[\frac{1}{\sqrt{2^n}} \sum_{z= 0}^{2^n-1} \bra z \right].
\end{equation}
Importantly, this Hamiltonian is invariant under the re-labelling of any pair of states, i.e. the under the permutation operator $\hat U_{j \leftrightarrow k}$ with
\begin{equation}
    \hat U_{j \leftrightarrow k} = \mathcal{I} - \ket{j}\bra{j}- \ket{k}\bra{k}+ \ket{k}\bra{j} + \ket{k}\bra{j}
\end{equation}
and so it holds that
\begin{equation}
        \hat H_{\rm G} = \hat U_{j \leftrightarrow k} \hat H_{\rm G} \hat U_{j \leftrightarrow k}
    \end{equation}
    for any permutation. Moreover, the initial state of QAOA bears the same property, as an eigenstate of any permutation operator one can write:
    \begin{equation}
        \ket{+}^{\otimes n} = \hat U_{j \leftrightarrow k} \ket{+}^{\otimes n}.
    \end{equation}
    Of course, a problem Hamiltonian is, in general, not invariant under any permutation. However, alongside a Grover-QAOA state (defined as in equation (\ref{QAOA_state}) using with $H_D = H_G$) its expectation is, since permutation operators may commute through the driver to annihilate or act as the identity on the initial state. So, for any permutation of a problem Hamiltonian $\hat H_{\rm P}$, given by $\hat P^\dagger \hat H_{\rm P} \hat P$ where 
    \begin{equation}    
    \hat P = \hat U_{j_m \leftrightarrow k_m}...\hat U_{j_2 \leftrightarrow k_2}\hat U_{j_1 \leftrightarrow k_1}
    \end{equation} 
    is a sequence of permutations, the QAOA problem expectation produced by such a permuted problem Hamiltonian can be shown to be equivalent to that of the unpermuted Hamiltonian, 
    \begin{equation}
        \langle \hat P^\dagger \hat H_{\textrm{P}} \hat P \rangle _{\vec \beta,\vec\gamma}=\langle \hat H_{\textrm{P}} \rangle _{\vec \beta,\vec\gamma},
    \end{equation}
as shown in appendix \ref{permutation_independence_of_expectation}.
The consequence of this symmetry is that any two problems which have the same distribution of energy levels will have identical expectation values under Grover driven QAOA states. 

\subsection{Problem Objective Functions as Random Variables}

The only property of the problem Hamiltonian that therefore affects the expectation is its spectrum. The spectrum of a problem Hamiltonian can be seen as sampling from some random distribution with probability density function $f(c)$---describing the relative likelihood that the random variable $C$, modelling a problem objective function, takes real value $c$---analogous to the density of states function of solid state physics. Our key result is that as this pdf can be, in general, independent of the size of a problem, angles can be calculated directly from the pdf, rather than for any model instance with given size. By computing angles for the large-$n$ limit of Grover Driven QAOA, one can determine the asymptotic performance of the algorithm for large system sizes without suffering from unfavorable scaling in the number of qubits. The computed angles are for ``typical'' instances. However, as the problem size increases, the law of large numbers rapidly reduces the sample-to-sample fluctuations.

Consider a problem's objective function that has been sorted from low to high, denoted by $C_{\rm sort}(z)$, such that 
\begin{equation}
    C_{\rm sort}(x) \geq C_{\rm sort}(y) \,\, \mathrm{if} \,\, x>y.
\end{equation}
    To this quantity we can associate a quantile distribution function (qdf)
\begin{equation}
    F^{-1}(p) = C_{\rm sort}(z)\,\, \mathrm{for}\,\, p = \frac{z}{N}
\end{equation}
denoting the value of C at its $p$th quantile. The inverse of the quantile distribution function is the cumulative distribution function (cdf) $F(c)$ with an associated pdf $f(c)$ in the continuum limit of large $n$
\begin{equation}
    F(c) = P(C < c) = \sum_{c' =-\infty}^c P(C = c') \approx \int_{-\infty}^c f(c')\,\mathrm{d}c'.
\end{equation}
The quantity on which the Grover driven QAOA expectation value depends is the Fourier transform of this pdf, known as the characteristic function $\Gamma$ which for a random variable $X$ is defined as the the expected value of $e^{i\gamma X}$ and so for the random variable $C$ modelling an ensemble of objective functions for a given problem:
\begin{equation}
    \Gamma(\gamma) = \mathrm{E}\left[e^{i\gamma C}\right].
\end{equation}
This quantity, associated with the random variable $C$ can be expressed in terms of the pdf $f(c)$, the cdf $F(c)$ and the qdf $F^{-1}(p)$ as: 
\begin{multline}
\Gamma(\gamma) = \int_{-\infty}^\infty f(c) e^{i\gamma c}\, \mathrm{d}c \\= \int_{-\infty}^\infty e^{i\gamma c}\, \mathrm{d}F(c) = \int_0^1 e^{i\gamma F^{-1}(p)} \mathrm{d}p
\end{multline}
and is related via a Wick rotation and scaling factor of the system size to the partition function of the model. 

A finite-size, $n$-qubit problem considered for QAOA is modelled here as taking $2^n = N$ samples from the ensemble random variable $C$ and as such the approximation made in this work is that the mean exponentiated objective function value can be replaced by the ensemble characteristic function $\Gamma$. That is, that 
\begin{equation}
\Gamma(\gamma) \approx  \sum_{z} \frac{e^{i\gamma C(z)}}{N}     \label{equation:gammasum} 
\end{equation}
with equality in the infinite $N$ limit. Due to the exponential growth of $N$ with the number of qubits, such an approximation rapidly improves with increasing system sizes. As such, it suffices to replace the finite sum in equation (\ref{equation:gammasum}) by the characteristic function of the continuous pdf---from which the problem is modelled to sample from---in calculations.

\section{Depth 1}

In this section we wish to obtain an expression for the expectation value of the depth $p=1$ Grover-QAOA problem Hamiltonian $E_1(\gamma, \beta)$ in which the problem is expressed only via a characteristic function $\Gamma$. We denote the mean value of a random variable $C$ as
\begin{equation}
    \bar C =  \int_{-\infty}^{\infty}cf(c)\,\mathrm{d}c \stackrel{n\to \infty}{=} \frac{1}{N}\sum_{z} C(z) = -i\Gamma'(0)
\end{equation}
which can in general be set to zero by adding a constant to the objective function. Such a shift corresponds only to a global phase of the effected problem Hamiltonian unitary $e^{-i\bar C \gamma} \mathcal{I}$. The expression derived, in full detail in appendix \ref{depth1_appendix_section} consists of factors for which equation \ref{equation:gammasum} can immediately be substituted, with, however, one instance of $C(z)$ as a multiplicative factor rather than in an exponent. To substitute this factor, one may differentiate under the sum/integral in the characteristic function as:
\begin{multline}
    \frac{1}{N}\sum_{z}C(z)\mathrm{e}^{i\gamma C(z)} = -\frac{i}{N}\frac{\mathrm{d}}{\mathrm{d}\gamma} \sum_{z}C(z)\mathrm{e}^{-i\gamma C(z)} \\
    = -i\frac{\mathrm{d}}{\mathrm{d}\gamma'} \Gamma(\gamma') \bigg\vert_{\gamma} = -i\Gamma'(\gamma).
\end{multline}
Substituting the characteristic function $\Gamma$ and its derivative above, one obtains the expression
\begin{equation}
E_1(\gamma, \beta) = \bar C(1 + BB^*\Gamma \Gamma^*) + 2\Im(B^*\Gamma^* \Gamma')
\label{p1_equation}
\end{equation}
in which the first term can be set to zero by zeroing the mean value of the problem.
\section{Depth 2}
The expectation value of the problem Hamiltonian for a $p = 2$ Grover-driven QAOA state depends on four parameters and can be expressed as: 
\begin{multline}
    E_2(\gamma_1,\gamma_2,\beta_1, \beta_2) = \bra{\gamma_1,\gamma_2,\beta_1, \beta_2} \hat H_P \ket{\gamma_1,\gamma_2,\beta_1, \beta_2}
\end{multline}
which, if one assumes that the mean of the distribution is zero, can be expanded in $10$ terms seen in appendix \ref{p2_calculation} with the help of computer algebra \cite{10.7717/peerj-cs.103}. Further simplification and substitution of the depth 1 expression yields:
\begin{multline}
    E_2(\gamma_1,\gamma_2,\beta_1, \beta_2) =\\ E_1(\gamma_1,\beta_1) + E_1(\gamma_1 + \gamma_2,\beta_2) 
    +\lvert{B(\beta_{1})\rvert}^2 \lvert{\Gamma}{\left(\gamma_{1} \right)}\rvert^2E_1(\gamma_2,\beta_2)
    \\
    +2\Im{B{\left(\beta_{1} \right)}^{*} B{\left(\beta_{2} \right)}^{*} \Gamma{\left(\gamma_{1} \right)}^{*} \Gamma{\left(\gamma_{2} \right)}^{*}  \Gamma '(\gamma_{1} + \gamma_{2})}
    \\
    +2\Im{B{\left(\beta_{1} \right)}  B{\left(\beta_{2} \right)}^{*} \Gamma{\left(\gamma_{1} \right)}  \Gamma ' {\left(\gamma_{2} \right)}\Gamma{\left(\gamma_{1} + \gamma_{2} \right)}^{*}} \label{depth2_final_expression}
\end{multline}
which can be numerically extremized on a classical computer to find optimal parameters given a characteristic function $\Gamma$.

\section{Depth $p$}
\label{p2_calculation}
For arbitrary depth $p$, we have the problem Hamiltonian expectation value of:
\begin{equation}
    E_p(\vec \gamma, \vec \beta)=     \bra{+}\prod_{j=-1} ^ {-p} \hat U_\mathrm{P}(\gamma_j)\hat U_\mathrm{D}(\beta_j) \hat H_{\rm P} \prod_{i=1} ^ {p}\hat U_\mathrm{D}(\beta_i)\hat U_\mathrm{P}(\gamma_i) \ket{+}    
\end{equation}
where we have introduced the convention that negative indices simply add a negative sign to the value, as: \begin{equation}
    \gamma_{-i} = -\gamma_i, \,\, \beta_{-i} = -\beta_i.
\end{equation} 
This can be expanded and expressed as a formula summing over $2^{2p}$ terms expressed in factors of the characteristic function $\Gamma$ and the driver dependent function $B$. The full derivation and explanation of notation used for which can be found in appendix \ref{arbitrary_depth_calculation} with the expression for the arbitrary depth expected value of the problem Hamiltonian taking the form:
\begin{widetext}
\begin{multline}
        E_p(\vec \gamma, \vec \beta) =  2\Im{\sum_{k_\mathrm{bra}< k_\mathrm{ket}=0}^{2^p-1}\prod_{P \in P_{\mathrm{bra}}} \Gamma\left(\sum_{i\in P} \gamma_i \right)\Gamma'\left(\sum_{i\in P_{\rm central}} \gamma_i \right)\prod_{P \in P_{\mathrm{ket}}} \Gamma\left(\sum_{i\in P} \gamma_i \right)\prod_{-j \vert k_{\mathrm{bra}}^j=1 \, \mathrm{or}\, j \vert k_{\mathrm{ket}}^j=1} B_j}.
\end{multline}
\end{widetext}

\section{Problem Distributions}

To find suitable problems for this algorithm, one requires problems for which the probability density function is known, but for which no efficient classical algorithms for finding states of extreme energy exist. The standard QAOA problems such Max-Cut, $k$-SAT and SK models approximately follow binomial and Gaussian distributions, but the presence of frustration in these problems result in distributions that, to the knowledge of the authors, do not have simple analytical forms. The random cost model and number partitioning problem do not feature the same frustration, with the random variables by which they are defined enter into the problem via single-qubit terms, with the number partitioning problem made non-trivial via a global constraint enforcing positivity of the cost function. As such, the distributions for the latter problems are known.

\subsection{The Random Cost Model}

\begin{figure}
    \centering
    \hspace{-0.0cm}
    \includegraphics[width = 9cm]{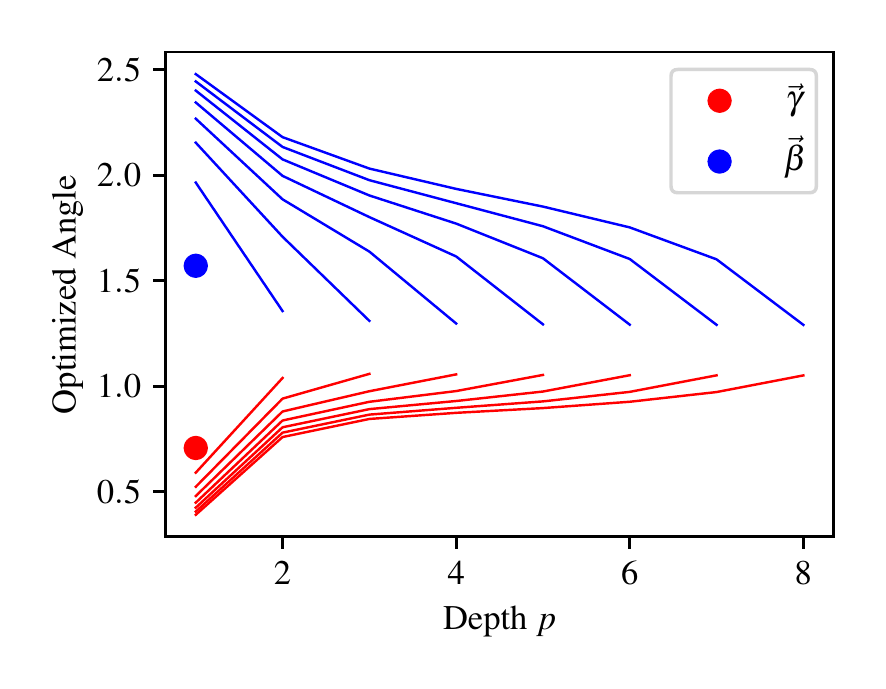}
    \vspace{-0.5cm}
    \caption{Angles for the optimized infinite-size RCM are seen for QAOA depths $1$ to $8$. As the depth $p$ increases, a similar pattern of optimal parameters for the limiting distribution of the random cost model emerges with the problem Hamiltonian angles monotonically increasing and the driver Hamiltonian angles monotonically decreasing. Such behavior invites comparison to adiabatic quantum computation, overlooking the non-zero start point of $\gamma$ and end point of $\beta$.}
    \label{fig:REM_parameters}
\end{figure}

\begin{figure}
    \centering
    \hspace{-0.0cm}
    \includegraphics[width = 9cm]{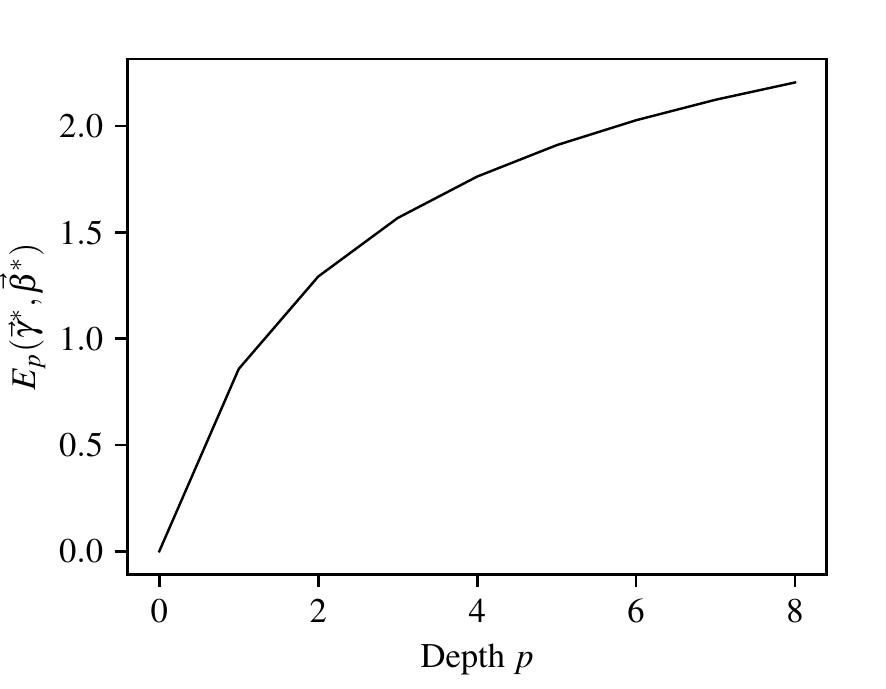}
    \vspace{-0.5cm}
    \caption{The expectation value of the QAOA state for the random cost model at optimal angles exhibits an increasing solution quality at a decreasing rate.}
    \label{fig:REM_expectation}
\end{figure}

The Random Cost Model (RCM) is a toy model used as a testing ground for techniques in the study of disordered systems as is the simplest model exhibiting a phase transition \cite{mezard2009information}. The model is not a description of any physical system, nor does it describe a hard problem. However, it does provide an example for which our procedure generates an concise analytical expression at low depth. The RCM is defined as one in which each energy level or objective function value samples from a normal distribution, so
\begin{equation}
    C \sim \mathrm{Normal}(\mu = 0,\sigma^2=  1), \,\, f(c) = \frac{1}{\sqrt{2\pi}} e^{-\frac{c^2}{2}}.
\end{equation}
When using the Grover driver, a problem with this pdf can also be obtained via $n$ independent spins with Hamiltonian
\begin{equation}
    \sum_i g_i Z_i = \sum_{z}\sum_i (-1)^{z_i}g_i \ket{z}\bra{z}
\end{equation}
in which each Pauli spin weight is set as $g_i \sim \mathrm{Normal}(\mu = 0,\sigma^2 = 1/ n)$ such that the resulting angles and energy expectations are independent of problem size. This Hamiltonian is, of course, structured and for another choice of driver could not be usefully modeled as a Gaussian pdf alone, the Grover driver, however, does not see this additional structure. This problem yields the desired pdf as a normalized Gaussian with which is associated a characteristic function via Fourier transform with 
\begin{equation}
     \Gamma(\gamma) = e^{-\frac{\gamma^2}{2}}.
 \end{equation} 
When substituted into equation \ref{p1_equation}, this characteristic function produces a $p = 1$ Grover-QAOA problem expectation value of 
 \begin{equation}
    E_1(\gamma,\beta) =  2 \gamma e^{- \gamma^{2}} \sin{\left(\beta \right)}
\end{equation}
for which minimization yields optimal angles of
 \begin{equation}
     \gamma = \frac{\sqrt2}{2}, \,\, \beta = \frac\pi{2}.
 \end{equation}
Angles for higher depth QAOA states can be found in figure \ref{fig:REM_parameters} and the associated expectation values in figure \ref{fig:REM_expectation}. The best known angles are determined via numerical optimization with BFGS \cite{wright1999numerical} on a wide range of start points, in which the consistent behavior of the angles with increasing depth implies that with increasing depth one tends towards a continuous terminal schedule.

\subsection{The Number Partitioning Problem}
\label{npp_section}
The Number Partitioning Problem (NPP) is that of finding a partition $z \in \{0...N -1\}$ of a set of $n$ positive real numbers $\{x_0,..., x_{n-1} \}$ indexed by the set $\{0,..., n-1 \}$ into two subsets with indices $\{i\,\, |\,\, z_i = 0\}$ and $\{i\,\, |\,\, z_i = 1\}$ such that the sum of the numbers indexed by each subset is equal. NPP has applications ranging from task scheduling in multi-core processing to choosing equally matched teams from a selection of players. It is an example of a problem for which classical heuristics such as simulated annealing do not perform well, notably, due to the similarity of the problem to totally unstructured cost problems \cite{mertens2000random}. The problem has seen study in the context of adiabatic quantum computing \cite{smelyanskiy2002dynamics} due to the relative ease of computing related properties in the large problem size limit, thus, allowing the calculation of the minimum energy gap and asymptotic time complexity in this context. In the large problem-size limit, adiabatic quantum computing returns a Grover-like quadratic speed up in run-time over random guessing for this problem. 

One may define the decision version of the NPP as to satisfy the equality:
\begin{equation}    
\sum_{i |z_i = 0} x_i = \sum_{i |z_i = 1} x_i.
\end{equation}
Which, for an approximate minimization version of the problem, is characterized by an objective function:
\begin{equation}
C(z) = \left\vert \sum_{i |z_i = 0} x_i - \sum_{i |z_i = 1} x_i\right\vert = \left\vert\sum_{i} (-1)^{z_i}x_i\right\vert.
\end{equation}
When considering this problem in the context of QAOA, or other quantum optimization contexts in which many-body interactions are nontrivial to implement, one can consider an equivalent minimization problem defined by the squared problem $C^2(z)$. To express this as an Ising spin glass Hamiltonian, one can first consider the signed single qubit problem Hamiltonian:
\begin{equation}
        \sum_i x_iZ_i = \sum_{z} \left(\sum_i{{(-1)^{z_i}}x_i}\right)\ket{z}\bra{z}
    \end{equation}
of which the square results in the problem Hamiltonian:
\begin{multline}
\sum_i x_iZ_i\sum_j x_j Z_j = \sum_i x_i^2 I + 2\sum_{i<j}x_ix_j Z_iZ_j\\
=\sum_{z, z'} \left(\sum_i{{(-1)^{z_i}}x_i}\right)\ket{z}\bra{z} \left(\sum_i{{(-1)^{z'_i}}x_i}\right)\ket{z'}\bra{z'}
\\
= \sum_z\sum_{i,j} (-1)^{z_i + z_j}x_i x_j\ket{z}\bra{z}\\
= \sum_z C^2(z)\ket{z}\bra{z} 
\end{multline}
which is a problem Hamiltonian encoding for the squared residue of of the number partitioning problem, as:
\begin{multline}
    C^2(z) = \Bigg\vert\sum_{i} (-1)^{z_i}x_i\Bigg\vert \cdot  \Bigg\vert\sum_{j} (-1)^{z_j}x_j \Bigg\vert
    \\
    = \sum_{i,j} (-1)^{z_i + z_j}x_i x_j .
\end{multline}
 Consequently, the NPP Hamiltonian is a all-to-all coupled Mattis-Type Ising spin glass in which couplings consist of the products of random variables. Now we must determine the pdf $f(c)$ and the associated partition function $\Gamma$ for this problem. For convenience, we consider the numbers to be partitioned to be sampled from distributions such that the resulting cost function distribution has unit mean. As such the numbers $x_i$ are allowed to take values between $0$ and $x_{\rm max} = 1 / \sqrt{3n}$ uniformly, i.e. $x_i \sim U(0,x_{\rm max}) \,\, \forall \,\,i\in[1...n]$. The probability density function for the individual variables is then defined as:
 \begin{equation}
     f(c) = \frac{1}{x_{\mathrm{max}}}\left(\Theta(c) - \Theta(c - x_\mathrm{max})\right)
 \end{equation}
 where $\Theta$ is the Heaviside step function. As such, the pdf for the distribution of residues defined as $C_{SQ}$, prior to squaring the cost function, is distributed as: 
 \begin{multline}
  C_{SQ} \sim \sum_i (-1)^{z_i} x_i = \sum_i (-1)^{z_i} U(0,x_{\rm max}) \\
  = \sum_i U(-x_{\rm max},x_{\rm max})
 \end{multline}
 where the factor $(-1)^{z_i}$ imparts a minus sign with probability $\frac12$ and results in a sum over uniform distributions over the symmetric interval $[-x_\mathrm{max}, x_\mathrm{max}]$. This sum of independent, identically distributed random variables, by the central limit theorem, approaches a Gaussian distribution in the large $n$ limit. So, in the large $n$ limit 
\begin{widetext}
    \begin{equation}
            C_{SQ}  \sim \mathrm{Normal}\left(\sum_i \mathrm{E}(U(-x_{\rm max}, x_{\rm max})), \sum_i \mathrm{Var}(U(-x_{\rm max},x_{\rm max}))\right)= \mathrm{Normal}\left(\mu = 0,\sigma^2 =1\right).
    \end{equation}
\end{widetext}
For the squared single-qubit cost function $C^2_{sq}(z) = C^2(z)$, one can use the standard result that the square of a Gaussian distributed random variable is the $\chi^2$ distribution with a single degree of freedom with pdf:
\begin{equation}
    f(c) =  \begin{cases} (2\pi ce^c)^{-\frac{1}{2}},&c>0;\\
    0,&{\text{otherwise}}.
    \end{cases}    
\end{equation}
The characteristic function of this pdf can readily be calculated as
\begin{equation}
    \Gamma(\gamma) = \sqrt{\frac{1}{1-2i\gamma}}.
\end{equation}

\begin{figure}
    \centering
    \includegraphics[width = 9cm]{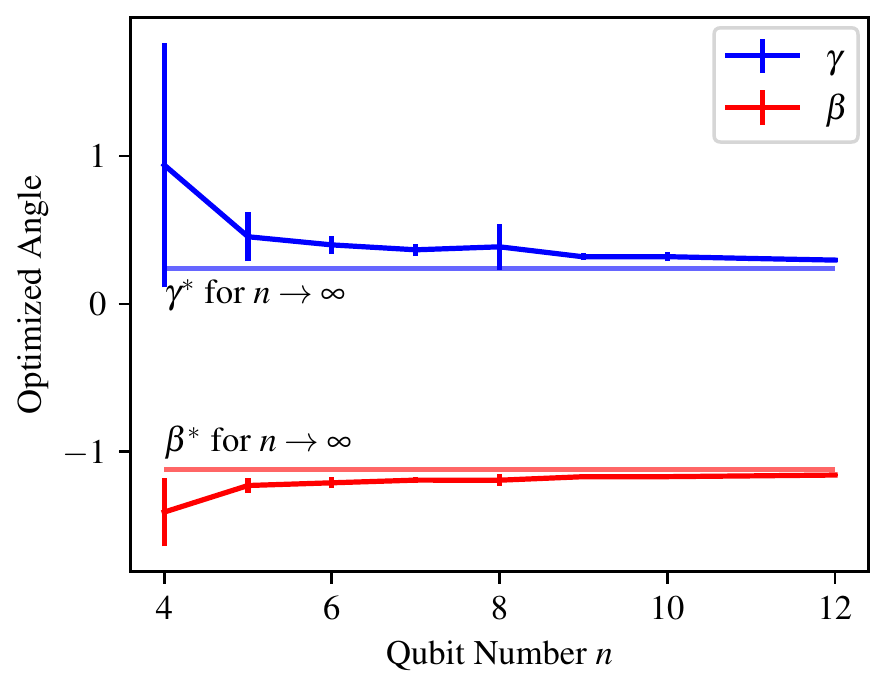}
    \vspace{-0.6cm}
    \caption{For $p=1$, optimal parameters of the NPP converge to the infinite size angle for increasing $n$. QAOA states for $30$ problem instances are optimized and angles averaged for each point.}
    \label{fig:p1_NPP}
\end{figure}

\begin{figure}
    \centering
    \includegraphics[width = 9cm]{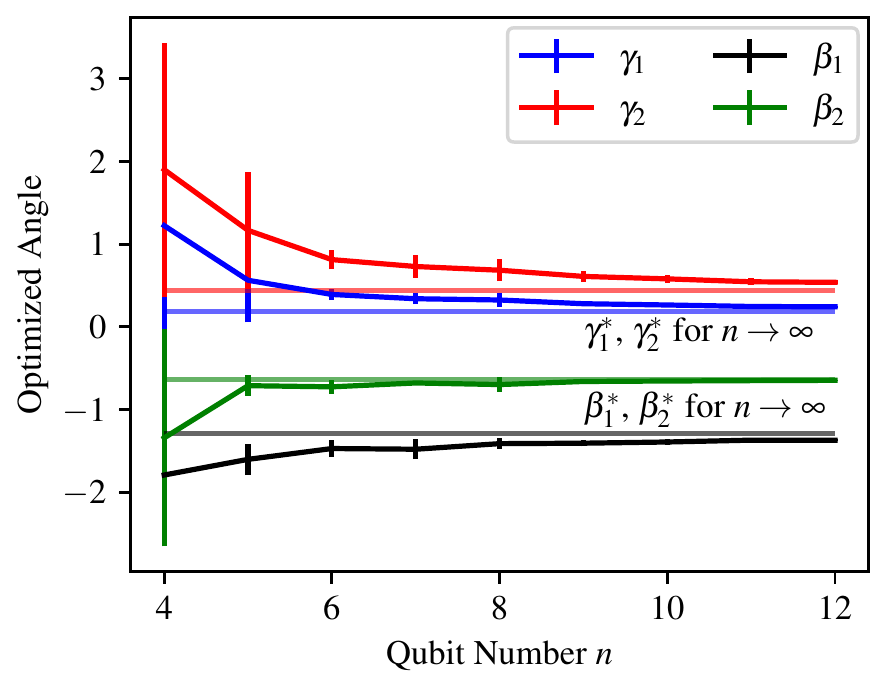}
    \vspace{-0.6cm}
    \caption{For $p=2$, optimal parameters of the NPP converge to the infinite size angle for increasing $n$. QAOA states for $30$ problem instances are optimized and angles averaged for each point.}
    \label{fig:p2_NPP}
\end{figure}

\begin{figure}
    \centering
    \hspace{-0cm}
    \includegraphics[width = 9cm]{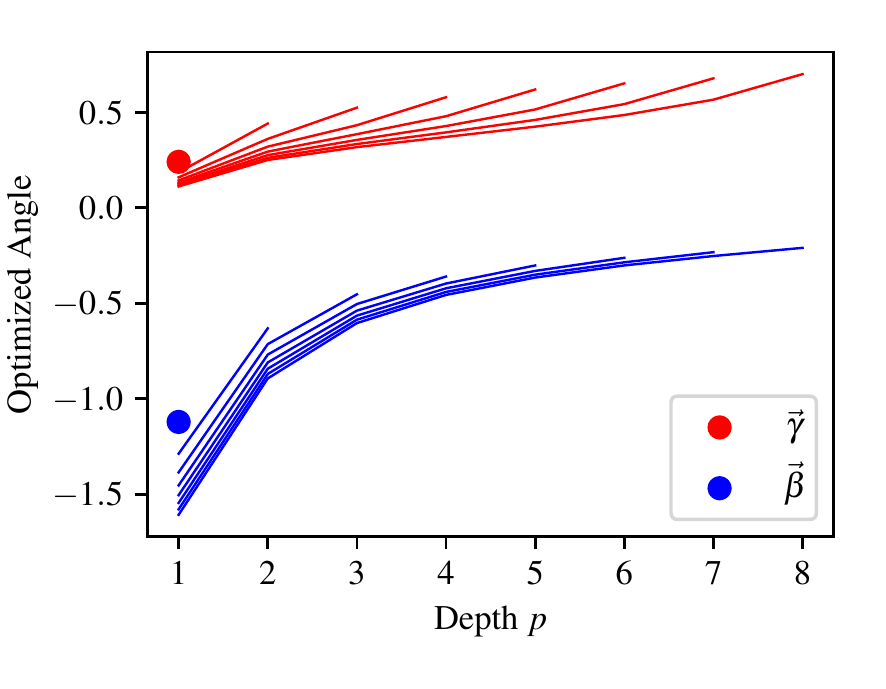}
    \vspace{-1 cm}
    \caption{Angles for the optimized infinite-size number partitioning problem are seen for QAOA depths $1$ to $8$. Optimal parameters follow annealing-reminiscent schedules and vary smoothly. Values are obtained here through the optimization of the formula derived in appendix \ref{arbitrary_depth_calculation} via the BFGS optimizer and a random start point strategy.}
    \label{fig:optimal_paraeters_NPP}
\end{figure}

\begin{figure}
    \centering
    \hspace{0 cm}
    \includegraphics[width = 9cm]{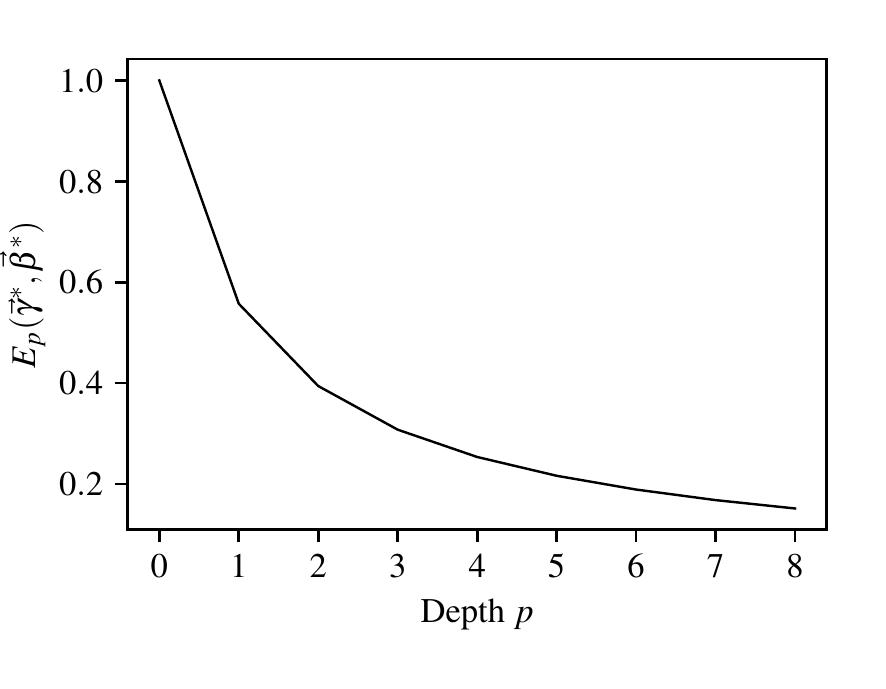}
    \vspace{-1cm}
    \caption{Optimizing the Grover-QAOA state for the number partitioning problem at increasing depth $p$ results in monotonically decreasing residue. The depth $0$ value of $1$ is added as a reference point attained by random sampling from the computational basis.}
    \label{fig:quality_NP}
\end{figure}

\begin{figure*}
\centering
   \hspace*{-1cm}\includegraphics{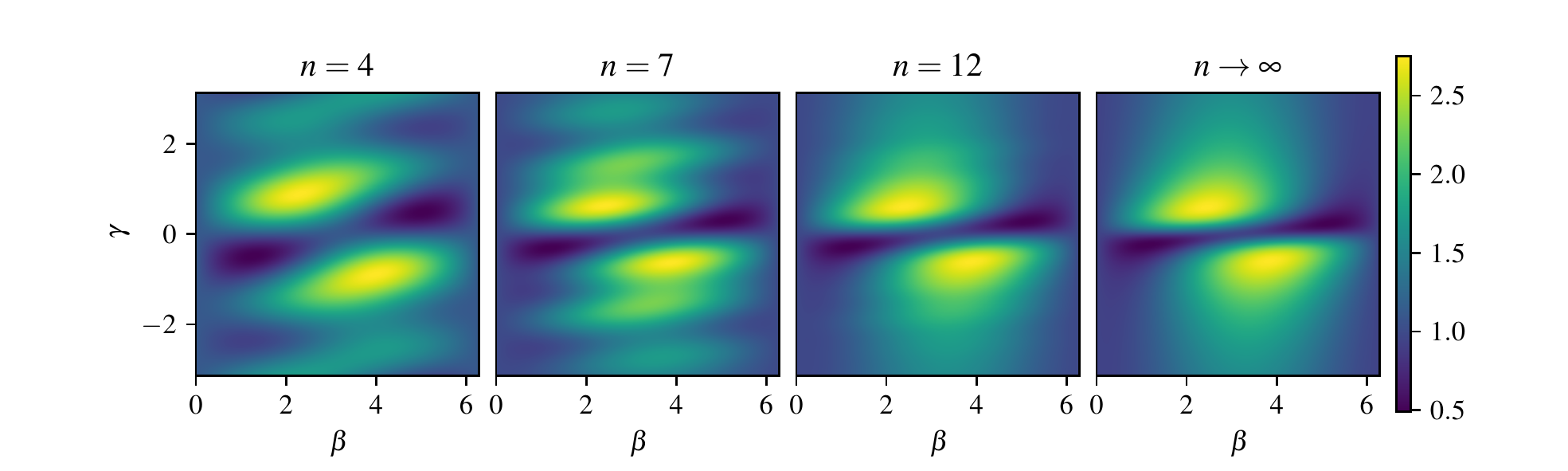}    
\vspace{-0.75cm}
    \caption{For $p = 1$ the number partitioning problem landscape can be seen to converge to that of the large-$n$ limit for $n=5,7,12$. Periodicity is always observed in $\beta$, but as $n$ increases, for low $n$, local maxima and minima resurge in $\gamma$ due to common multiples in problem energy eigenvalues. For large $n$, increasing $\gamma$ past a region near zero results in the expectation monotonically returning to the mean value of $1$.}
    \label{fig:heat_convergence}
\end{figure*}

For $p = 1$, one can numerically optimize the expectation value $E_1(\gamma,\beta)$ to find minimizing angles of 
\begin{equation}
\gamma^* = 0.241, \,\, \beta^* = 5.162 
\end{equation} 
to three decimal places. These angles can be seen in figure \ref{fig:p1_NPP} in which the value of $\beta$ has been shifted by $2\pi$ for convenience. Such angles yield an expected value of $E_1(\vec\gamma^*,\vec\beta^*) = 0.557$. The expectation value of the problem Hamiltonian for this characteristic function can be minimized to obtain angles that are thought to be optimal the best known angles can be seen in figure \ref{fig:optimal_paraeters_NPP}. The associated qualities of the solutions attained by these angles seen in figure \ref{fig:quality_NP}. To demonstrate that these angles are appropriate for finite-size instances of the number partitioning problem, we plot in figure \ref{fig:p1_NPP},\ref{fig:p2_NPP} the average result when optimizing finite-size instances at increasing size for depth $1$, $2$ respectively. This demonstrates that with increasing problem size, the optimal angles for finite-size instances rapidly approaches those of the analytically derived, ensemble average optimal angles. Finally, figure \ref{fig:heat_convergence} demonstrates the convergence of the landscape defined by the problem Hamiltonian expectation value of the QAOA state at a given point in the QAOA parameter space. As the number of qubits in the problem becomes larger, the QAOA solution quality landscape becomes indistinguishable to the analytically derived version.



\section{Compiling the Grover Driver}
Compilation of the Grover mixing oracle can be attained with or without ancillary qubits. Without ancillary qubits the operation can be performed using an $n-1$-controlled-$Z$ rotation. Such a gate can be compiled to two $n$-qubit Toffoli gates alongside three two-qubit gates \cite{barenco1995elementary}. These Toffoli gates can in turn be compiled using a number of two-qubit gates and depth quadratic and linear, respectively, in the number of qubits \cite{saeedi2013linear, he2017decompositions}.

\section{Conclusion}

The power of QAOA as a heuristic for solving problems in combinatorial optimization is not yet certain and derives from some combination of Grover-like speedup alongside performance attributable to the synergy of problem and driver used. In this work we demonstrate a  method to determine likely optimal QAOA angles in the case of a driver with average-case synergy, in which the problem structure cannot be exploited. For such problem-driver combinations, concentrations of optimal angles occur for increasing problem sizes and tend to a limit calculable via the pdf of the problem. It is expected that barren plateaus, limiting the ability of an optimizer to find good angles for QAOA, will affect problem driver combinations in which synergy is not present (for example, problems with high-order terms and single-qubit drivers). The problem of barren plateaus---in essence the No Free Lunch Theorem of quantum optimization \cite{wolpert1997no}---necessitates that one finds heuristics for which drivers, angles to use for a given problem instance. For which, performance of a driver in comparison to the Grover version is instructive as to whether a strategy exploits problem structure and to what degree. The strategy of optimizing QAOA angles for statistically defined ensembles of problems, rather than instances thereof, comes with the added advantage of smoother QAOA objective function landscapes with fewer local extrema, being the average over instance landscapes, as evidenced by figure \ref{fig:heat_convergence}.

This work originated from asking the question as to whether a single-qubit-$X$ driver should be outperform the Grover driver as an ansatz for the infinite range models in the work of  Farhi et. al \cite{farhi2019quantum}. While this work has not directly compared the performance of Grover QAOA on the SK model, a future direction could compare this, via application of methods applied to infinite-range SK models, to the number partitioning problem, or by finding or approximating an expression for the limiting-size pdf of an SK-type model. Future work could also investigate the performance of subspace-type drivers e.g. $XY$ type in problems of hard and soft constraints of limiting size \cite{hadfield2017quantum}, in which the form of the drivers on the valid-solution subspace differs from that of Grover drivers or standard single-qubit mixers. Work could also determine the dependence of QAOA performance on the shape of the distribution of the unstructured random variable on which it acts, as has been considered in the context of non-adiabatic annealing  \cite{yan2022analytical}. Various works have provided analytical results for the use of Grover-type problems or mixing operators to solve search problems in the context of QAOA, quantum random walks and adiabatic quantum computing \cite{morley2019quantum, jiang2017near, roland2002quantum, akshay2021parameter}, with this work demonstrating similar analytical procedures can be applied to find optimal angles for Grover driven QAOA with non-oracle problems. The work in this paper could also be applied to other contexts, such as that of quantum random walks, in which fully connected graphs have been previously been considered for numerical studies on finding ground states of spin glasses \cite{callison2019finding}.

\section*{Acknowledgments}

We acknowledge the support of Mercedes-Benz AG and the German Federal Ministry of Education and Research in the funding program “quantum technologies – from basic research to market” (contract number 13N15582). We thank Dmitry Bagrets for bringing the number partitioning problem to our attention as a hard problem for which the probability density function is known.






\bibliography{ref} 

\begin{thebibliography}{38}%
\makeatletter
\providecommand \@ifxundefined [1]{%
 \@ifx{#1\undefined}
}%
\providecommand \@ifnum [1]{%
 \ifnum #1\expandafter \@firstoftwo
 \else \expandafter \@secondoftwo
 \fi
}%
\providecommand \@ifx [1]{%
 \ifx #1\expandafter \@firstoftwo
 \else \expandafter \@secondoftwo
 \fi
}%
\providecommand \natexlab [1]{#1}%
\providecommand \enquote  [1]{``#1''}%
\providecommand \bibnamefont  [1]{#1}%
\providecommand \bibfnamefont [1]{#1}%
\providecommand \citenamefont [1]{#1}%
\providecommand \href@noop [0]{\@secondoftwo}%
\providecommand \href [0]{\begingroup \@sanitize@url \@href}%
\providecommand \@href[1]{\@@startlink{#1}\@@href}%
\providecommand \@@href[1]{\endgroup#1\@@endlink}%
\providecommand \@sanitize@url [0]{\catcode `\\12\catcode `\$12\catcode
  `\&12\catcode `\#12\catcode `\^12\catcode `\_12\catcode `\%12\relax}%
\providecommand \@@startlink[1]{}%
\providecommand \@@endlink[0]{}%
\providecommand \url  [0]{\begingroup\@sanitize@url \@url }%
\providecommand \@url [1]{\endgroup\@href {#1}{\urlprefix }}%
\providecommand \urlprefix  [0]{URL }%
\providecommand \Eprint [0]{\href }%
\providecommand \doibase [0]{http://dx.doi.org/}%
\providecommand \selectlanguage [0]{\@gobble}%
\providecommand \bibinfo  [0]{\@secondoftwo}%
\providecommand \bibfield  [0]{\@secondoftwo}%
\providecommand \translation [1]{[#1]}%
\providecommand \BibitemOpen [0]{}%
\providecommand \bibitemStop [0]{}%
\providecommand \bibitemNoStop [0]{.\EOS\space}%
\providecommand \EOS [0]{\spacefactor3000\relax}%
\providecommand \BibitemShut  [1]{\csname bibitem#1\endcsname}%
\let\auto@bib@innerbib\@empty
\bibitem [{\citenamefont {Preskill}(2018)}]{preskill2018quantum}%
  \BibitemOpen
  \bibfield  {author} {\bibinfo {author} {\bibfnamefont {J.}~\bibnamefont
  {Preskill}},\ }\href@noop {} {\bibfield  {journal} {\bibinfo  {journal}
  {Quantum}\ }\textbf {\bibinfo {volume} {2}},\ \bibinfo {pages} {79} (\bibinfo
  {year} {2018})}\BibitemShut {NoStop}%
\bibitem [{\citenamefont {Farhi}\ \emph {et~al.}(2014)\citenamefont {Farhi},
  \citenamefont {Goldstone},\ and\ \citenamefont {Gutmann}}]{farhi2014quantum}%
  \BibitemOpen
  \bibfield  {author} {\bibinfo {author} {\bibfnamefont {E.}~\bibnamefont
  {Farhi}}, \bibinfo {author} {\bibfnamefont {J.}~\bibnamefont {Goldstone}}, \
  and\ \bibinfo {author} {\bibfnamefont {S.}~\bibnamefont {Gutmann}},\
  }\href@noop {} {\bibfield  {journal} {\bibinfo  {journal} {arXiv preprint
  arXiv:1411.4028}\ } (\bibinfo {year} {2014})}\BibitemShut {NoStop}%
\bibitem [{\citenamefont {Lloyd}(2018)}]{lloyd2018quantum}%
  \BibitemOpen
  \bibfield  {author} {\bibinfo {author} {\bibfnamefont {S.}~\bibnamefont
  {Lloyd}},\ }\href@noop {} {\bibfield  {journal} {\bibinfo  {journal} {arXiv
  preprint arXiv:1812.11075}\ } (\bibinfo {year} {2018})}\BibitemShut {NoStop}%
\bibitem [{\citenamefont {Morales}\ \emph {et~al.}(2020)\citenamefont
  {Morales}, \citenamefont {Biamonte},\ and\ \citenamefont
  {Zimbor{\'a}s}}]{morales2020universality}%
  \BibitemOpen
  \bibfield  {author} {\bibinfo {author} {\bibfnamefont {M.~E.}\ \bibnamefont
  {Morales}}, \bibinfo {author} {\bibfnamefont {J.~D.}\ \bibnamefont
  {Biamonte}}, \ and\ \bibinfo {author} {\bibfnamefont {Z.}~\bibnamefont
  {Zimbor{\'a}s}},\ }\href@noop {} {\bibfield  {journal} {\bibinfo  {journal}
  {Quantum Information Processing}\ }\textbf {\bibinfo {volume} {19}},\
  \bibinfo {pages} {1} (\bibinfo {year} {2020})}\BibitemShut {NoStop}%
\bibitem [{\citenamefont {Jiang}\ \emph
  {et~al.}(2017{\natexlab{a}})\citenamefont {Jiang}, \citenamefont {Rieffel},\
  and\ \citenamefont {Wang}}]{GroverQAOANASA}%
  \BibitemOpen
  \bibfield  {author} {\bibinfo {author} {\bibfnamefont {Z.}~\bibnamefont
  {Jiang}}, \bibinfo {author} {\bibfnamefont {E.~G.}\ \bibnamefont {Rieffel}},
  \ and\ \bibinfo {author} {\bibfnamefont {Z.}~\bibnamefont {Wang}},\ }\href
  {\doibase 10.1103/physreva.95.062317} {\bibfield  {journal} {\bibinfo
  {journal} {Physical Review A}\ }\textbf {\bibinfo {volume} {95}} (\bibinfo
  {year} {2017}{\natexlab{a}}),\ 10.1103/physreva.95.062317}\BibitemShut
  {NoStop}%
\bibitem [{\citenamefont {Akshay}\ \emph {et~al.}(2021)\citenamefont {Akshay},
  \citenamefont {Rabinovich}, \citenamefont {Campos},\ and\ \citenamefont
  {Biamonte}}]{akshay2021parameter}%
  \BibitemOpen
  \bibfield  {author} {\bibinfo {author} {\bibfnamefont {V.}~\bibnamefont
  {Akshay}}, \bibinfo {author} {\bibfnamefont {D.}~\bibnamefont {Rabinovich}},
  \bibinfo {author} {\bibfnamefont {E.}~\bibnamefont {Campos}}, \ and\ \bibinfo
  {author} {\bibfnamefont {J.}~\bibnamefont {Biamonte}},\ }\href@noop {}
  {\bibfield  {journal} {\bibinfo  {journal} {Physical Review A}\ }\textbf
  {\bibinfo {volume} {104}},\ \bibinfo {pages} {L010401} (\bibinfo {year}
  {2021})}\BibitemShut {NoStop}%
\bibitem [{\citenamefont {Rabinovich}\ \emph {et~al.}(2021)\citenamefont
  {Rabinovich}, \citenamefont {Sengupta}, \citenamefont {Campos}, \citenamefont
  {Akshay},\ and\ \citenamefont {Biamonte}}]{rabinovich2021progress}%
  \BibitemOpen
  \bibfield  {author} {\bibinfo {author} {\bibfnamefont {D.}~\bibnamefont
  {Rabinovich}}, \bibinfo {author} {\bibfnamefont {R.}~\bibnamefont
  {Sengupta}}, \bibinfo {author} {\bibfnamefont {E.}~\bibnamefont {Campos}},
  \bibinfo {author} {\bibfnamefont {V.}~\bibnamefont {Akshay}}, \ and\ \bibinfo
  {author} {\bibfnamefont {J.}~\bibnamefont {Biamonte}},\ }\href@noop {}
  {\bibfield  {journal} {\bibinfo  {journal} {arXiv preprint arXiv:2109.11566}\
  } (\bibinfo {year} {2021})}\BibitemShut {NoStop}%
\bibitem [{\citenamefont {Streif}\ and\ \citenamefont
  {Leib}(2020)}]{streif2020training}%
  \BibitemOpen
  \bibfield  {author} {\bibinfo {author} {\bibfnamefont {M.}~\bibnamefont
  {Streif}}\ and\ \bibinfo {author} {\bibfnamefont {M.}~\bibnamefont {Leib}},\
  }\href@noop {} {\bibfield  {journal} {\bibinfo  {journal} {Quantum Science
  and Technology}\ }\textbf {\bibinfo {volume} {5}},\ \bibinfo {pages} {034008}
  (\bibinfo {year} {2020})}\BibitemShut {NoStop}%
\bibitem [{\citenamefont {Farhi}\ \emph {et~al.}(2019)\citenamefont {Farhi},
  \citenamefont {Goldstone}, \citenamefont {Gutmann},\ and\ \citenamefont
  {Zhou}}]{farhi2019quantum}%
  \BibitemOpen
  \bibfield  {author} {\bibinfo {author} {\bibfnamefont {E.}~\bibnamefont
  {Farhi}}, \bibinfo {author} {\bibfnamefont {J.}~\bibnamefont {Goldstone}},
  \bibinfo {author} {\bibfnamefont {S.}~\bibnamefont {Gutmann}}, \ and\
  \bibinfo {author} {\bibfnamefont {L.}~\bibnamefont {Zhou}},\ }\href@noop {}
  {\bibfield  {journal} {\bibinfo  {journal} {arXiv preprint arXiv:1910.08187}\
  } (\bibinfo {year} {2019})}\BibitemShut {NoStop}%
\bibitem [{\citenamefont {Basso}\ \emph {et~al.}(2021)\citenamefont {Basso},
  \citenamefont {Farhi}, \citenamefont {Marwaha}, \citenamefont {Villalonga},\
  and\ \citenamefont {Zhou}}]{basso2021quantum}%
  \BibitemOpen
  \bibfield  {author} {\bibinfo {author} {\bibfnamefont {J.}~\bibnamefont
  {Basso}}, \bibinfo {author} {\bibfnamefont {E.}~\bibnamefont {Farhi}},
  \bibinfo {author} {\bibfnamefont {K.}~\bibnamefont {Marwaha}}, \bibinfo
  {author} {\bibfnamefont {B.}~\bibnamefont {Villalonga}}, \ and\ \bibinfo
  {author} {\bibfnamefont {L.}~\bibnamefont {Zhou}},\ }\href@noop {} {\bibfield
   {journal} {\bibinfo  {journal} {arXiv preprint arXiv:2110.14206}\ }
  (\bibinfo {year} {2021})}\BibitemShut {NoStop}%
\bibitem [{\citenamefont {Boulebnane}\ and\ \citenamefont
  {Montanaro}(2021)}]{boulebnane2021predicting}%
  \BibitemOpen
  \bibfield  {author} {\bibinfo {author} {\bibfnamefont {S.}~\bibnamefont
  {Boulebnane}}\ and\ \bibinfo {author} {\bibfnamefont {A.}~\bibnamefont
  {Montanaro}},\ }\href@noop {} {\bibfield  {journal} {\bibinfo  {journal}
  {arXiv preprint arXiv:2110.10685}\ } (\bibinfo {year} {2021})}\BibitemShut
  {NoStop}%
\bibitem [{\citenamefont {Claes}\ and\ \citenamefont {van
  Dam}(2021)}]{claes2021instance}%
  \BibitemOpen
  \bibfield  {author} {\bibinfo {author} {\bibfnamefont {J.}~\bibnamefont
  {Claes}}\ and\ \bibinfo {author} {\bibfnamefont {W.}~\bibnamefont {van
  Dam}},\ }\href@noop {} {\bibfield  {journal} {\bibinfo  {journal} {Quantum}\
  }\textbf {\bibinfo {volume} {5}},\ \bibinfo {pages} {542} (\bibinfo {year}
  {2021})}\BibitemShut {NoStop}%
\bibitem [{\citenamefont {McClean}\ \emph {et~al.}(2018)\citenamefont
  {McClean}, \citenamefont {Boixo}, \citenamefont {Smelyanskiy}, \citenamefont
  {Babbush},\ and\ \citenamefont {Neven}}]{mcclean2018barren}%
  \BibitemOpen
  \bibfield  {author} {\bibinfo {author} {\bibfnamefont {J.~R.}\ \bibnamefont
  {McClean}}, \bibinfo {author} {\bibfnamefont {S.}~\bibnamefont {Boixo}},
  \bibinfo {author} {\bibfnamefont {V.~N.}\ \bibnamefont {Smelyanskiy}},
  \bibinfo {author} {\bibfnamefont {R.}~\bibnamefont {Babbush}}, \ and\
  \bibinfo {author} {\bibfnamefont {H.}~\bibnamefont {Neven}},\ }\href@noop {}
  {\bibfield  {journal} {\bibinfo  {journal} {Nature communications}\ }\textbf
  {\bibinfo {volume} {9}},\ \bibinfo {pages} {1} (\bibinfo {year}
  {2018})}\BibitemShut {NoStop}%
\bibitem [{\citenamefont {Cerezo de~la Roca}\ \emph {et~al.}(2021)\citenamefont
  {Cerezo de~la Roca}, \citenamefont {Sone}, \citenamefont {Volkoff},
  \citenamefont {Cincio},\ and\ \citenamefont {Coles}}]{cerezo2021cost}%
  \BibitemOpen
  \bibfield  {author} {\bibinfo {author} {\bibfnamefont {M.~V.~S.}\
  \bibnamefont {Cerezo de~la Roca}}, \bibinfo {author} {\bibfnamefont
  {A.}~\bibnamefont {Sone}}, \bibinfo {author} {\bibfnamefont {T.~J.}\
  \bibnamefont {Volkoff}}, \bibinfo {author} {\bibfnamefont {L.}~\bibnamefont
  {Cincio}}, \ and\ \bibinfo {author} {\bibfnamefont {P.~J.}\ \bibnamefont
  {Coles}},\ }\href@noop {} {\bibfield  {journal} {\bibinfo  {journal} {Nature
  Communications}\ }\textbf {\bibinfo {volume} {12}} (\bibinfo {year}
  {2021})}\BibitemShut {NoStop}%
\bibitem [{\citenamefont {Zalka}(1999)}]{zalka1999grover}%
  \BibitemOpen
  \bibfield  {author} {\bibinfo {author} {\bibfnamefont {C.}~\bibnamefont
  {Zalka}},\ }\href@noop {} {\bibfield  {journal} {\bibinfo  {journal}
  {Physical Review A}\ }\textbf {\bibinfo {volume} {60}},\ \bibinfo {pages}
  {2746} (\bibinfo {year} {1999})}\BibitemShut {NoStop}%
\bibitem [{\citenamefont {B{\"a}rtschi}\ and\ \citenamefont
  {Eidenbenz}(2020)}]{bartschi2020grover}%
  \BibitemOpen
  \bibfield  {author} {\bibinfo {author} {\bibfnamefont {A.}~\bibnamefont
  {B{\"a}rtschi}}\ and\ \bibinfo {author} {\bibfnamefont {S.}~\bibnamefont
  {Eidenbenz}},\ }in\ \href@noop {} {\emph {\bibinfo {booktitle} {2020 IEEE
  International Conference on Quantum Computing and Engineering (QCE)}}}\
  (\bibinfo {organization} {IEEE},\ \bibinfo {year} {2020})\ pp.\ \bibinfo
  {pages} {72--82}\BibitemShut {NoStop}%
\bibitem [{\citenamefont {Golden}\ \emph {et~al.}(2021)\citenamefont {Golden},
  \citenamefont {B{\"a}rtschi}, \citenamefont {O’Malley},\ and\ \citenamefont
  {Eidenbenz}}]{golden2021threshold}%
  \BibitemOpen
  \bibfield  {author} {\bibinfo {author} {\bibfnamefont {J.}~\bibnamefont
  {Golden}}, \bibinfo {author} {\bibfnamefont {A.}~\bibnamefont
  {B{\"a}rtschi}}, \bibinfo {author} {\bibfnamefont {D.}~\bibnamefont
  {O’Malley}}, \ and\ \bibinfo {author} {\bibfnamefont {S.}~\bibnamefont
  {Eidenbenz}},\ }in\ \href@noop {} {\emph {\bibinfo {booktitle} {2021 IEEE
  International Conference on Quantum Computing and Engineering (QCE)}}}\
  (\bibinfo {organization} {IEEE},\ \bibinfo {year} {2021})\ pp.\ \bibinfo
  {pages} {137--147}\BibitemShut {NoStop}%
\bibitem [{\citenamefont {Golden}\ \emph {et~al.}(2022)\citenamefont {Golden},
  \citenamefont {B{\"a}rtschi}, \citenamefont {Eidenbenz},\ and\ \citenamefont
  {O'Malley}}]{golden2022evidence}%
  \BibitemOpen
  \bibfield  {author} {\bibinfo {author} {\bibfnamefont {J.}~\bibnamefont
  {Golden}}, \bibinfo {author} {\bibfnamefont {A.}~\bibnamefont
  {B{\"a}rtschi}}, \bibinfo {author} {\bibfnamefont {S.}~\bibnamefont
  {Eidenbenz}}, \ and\ \bibinfo {author} {\bibfnamefont {D.}~\bibnamefont
  {O'Malley}},\ }\href@noop {} {\bibfield  {journal} {\bibinfo  {journal}
  {arXiv preprint arXiv:2202.00648}\ } (\bibinfo {year} {2022})}\BibitemShut
  {NoStop}%
\bibitem [{\citenamefont {Crooks}(2018)}]{crooks2018performance}%
  \BibitemOpen
  \bibfield  {author} {\bibinfo {author} {\bibfnamefont {G.~E.}\ \bibnamefont
  {Crooks}},\ }\href@noop {} {\bibfield  {journal} {\bibinfo  {journal} {arXiv
  preprint arXiv:1811.08419}\ } (\bibinfo {year} {2018})}\BibitemShut {NoStop}%
\bibitem [{\citenamefont {Mezard}\ and\ \citenamefont
  {Montanari}(2009)}]{mezard2009information}%
  \BibitemOpen
  \bibfield  {author} {\bibinfo {author} {\bibfnamefont {M.}~\bibnamefont
  {Mezard}}\ and\ \bibinfo {author} {\bibfnamefont {A.}~\bibnamefont
  {Montanari}},\ }\href@noop {} {\emph {\bibinfo {title} {Information, physics,
  and computation}}}\ (\bibinfo  {publisher} {Oxford University Press},\
  \bibinfo {year} {2009})\BibitemShut {NoStop}%
\bibitem [{\citenamefont {Akshay}\ \emph {et~al.}(2020)\citenamefont {Akshay},
  \citenamefont {Philathong}, \citenamefont {Morales},\ and\ \citenamefont
  {Biamonte}}]{akshay2020reachability}%
  \BibitemOpen
  \bibfield  {author} {\bibinfo {author} {\bibfnamefont {V.}~\bibnamefont
  {Akshay}}, \bibinfo {author} {\bibfnamefont {H.}~\bibnamefont {Philathong}},
  \bibinfo {author} {\bibfnamefont {M.~E.}\ \bibnamefont {Morales}}, \ and\
  \bibinfo {author} {\bibfnamefont {J.~D.}\ \bibnamefont {Biamonte}},\
  }\href@noop {} {\bibfield  {journal} {\bibinfo  {journal} {Physical review
  letters}\ }\textbf {\bibinfo {volume} {124}},\ \bibinfo {pages} {090504}
  (\bibinfo {year} {2020})}\BibitemShut {NoStop}%
\bibitem [{\citenamefont {Hadfield}(2018)}]{hadfield2018quantum}%
  \BibitemOpen
  \bibfield  {author} {\bibinfo {author} {\bibfnamefont {S.~A.}\ \bibnamefont
  {Hadfield}},\ }\href@noop {} {\emph {\bibinfo {title} {Quantum algorithms for
  scientific computing and approximate optimization}}}\ (\bibinfo  {publisher}
  {Columbia University},\ \bibinfo {year} {2018})\BibitemShut {NoStop}%
\bibitem [{\citenamefont {Zhu}\ \emph {et~al.}(2020)\citenamefont {Zhu},
  \citenamefont {Tang}, \citenamefont {Barron}, \citenamefont
  {Calderon-Vargas}, \citenamefont {Mayhall}, \citenamefont {Barnes},\ and\
  \citenamefont {Economou}}]{zhu2020adaptive}%
  \BibitemOpen
  \bibfield  {author} {\bibinfo {author} {\bibfnamefont {L.}~\bibnamefont
  {Zhu}}, \bibinfo {author} {\bibfnamefont {H.~L.}\ \bibnamefont {Tang}},
  \bibinfo {author} {\bibfnamefont {G.~S.}\ \bibnamefont {Barron}}, \bibinfo
  {author} {\bibfnamefont {F.}~\bibnamefont {Calderon-Vargas}}, \bibinfo
  {author} {\bibfnamefont {N.~J.}\ \bibnamefont {Mayhall}}, \bibinfo {author}
  {\bibfnamefont {E.}~\bibnamefont {Barnes}}, \ and\ \bibinfo {author}
  {\bibfnamefont {S.~E.}\ \bibnamefont {Economou}},\ }\href@noop {} {\bibfield
  {journal} {\bibinfo  {journal} {arXiv preprint arXiv:2005.10258}\ } (\bibinfo
  {year} {2020})}\BibitemShut {NoStop}%
\bibitem [{\citenamefont {Chen}\ \emph {et~al.}(2022)\citenamefont {Chen},
  \citenamefont {Zhu}, \citenamefont {Liu}, \citenamefont {Mayhall},
  \citenamefont {Barnes},\ and\ \citenamefont {Economou}}]{chen2022much}%
  \BibitemOpen
  \bibfield  {author} {\bibinfo {author} {\bibfnamefont {Y.}~\bibnamefont
  {Chen}}, \bibinfo {author} {\bibfnamefont {L.}~\bibnamefont {Zhu}}, \bibinfo
  {author} {\bibfnamefont {C.}~\bibnamefont {Liu}}, \bibinfo {author}
  {\bibfnamefont {N.~J.}\ \bibnamefont {Mayhall}}, \bibinfo {author}
  {\bibfnamefont {E.}~\bibnamefont {Barnes}}, \ and\ \bibinfo {author}
  {\bibfnamefont {S.~E.}\ \bibnamefont {Economou}},\ }\href@noop {} {\bibfield
  {journal} {\bibinfo  {journal} {arXiv preprint arXiv:2205.12283}\ } (\bibinfo
  {year} {2022})}\BibitemShut {NoStop}%
\bibitem [{\citenamefont {Hadfield}\ \emph {et~al.}(2017)\citenamefont
  {Hadfield}, \citenamefont {Wang}, \citenamefont {Rieffel}, \citenamefont
  {O'Gorman}, \citenamefont {Venturelli},\ and\ \citenamefont
  {Biswas}}]{hadfield2017quantum}%
  \BibitemOpen
  \bibfield  {author} {\bibinfo {author} {\bibfnamefont {S.}~\bibnamefont
  {Hadfield}}, \bibinfo {author} {\bibfnamefont {Z.}~\bibnamefont {Wang}},
  \bibinfo {author} {\bibfnamefont {E.~G.}\ \bibnamefont {Rieffel}}, \bibinfo
  {author} {\bibfnamefont {B.}~\bibnamefont {O'Gorman}}, \bibinfo {author}
  {\bibfnamefont {D.}~\bibnamefont {Venturelli}}, \ and\ \bibinfo {author}
  {\bibfnamefont {R.}~\bibnamefont {Biswas}},\ }in\ \href@noop {} {\emph
  {\bibinfo {booktitle} {Proceedings of the Second International Workshop on
  Post Moores Era Supercomputing}}}\ (\bibinfo {year} {2017})\ pp.\ \bibinfo
  {pages} {15--21}\BibitemShut {NoStop}%
\bibitem [{\citenamefont {Meurer}\ \emph {et~al.}(2017)\citenamefont {Meurer},
  \citenamefont {Smith}, \citenamefont {Paprocki}, \citenamefont
  {\v{C}ert\'{i}k}, \citenamefont {Kirpichev}, \citenamefont {Rocklin},
  \citenamefont {Kumar}, \citenamefont {Ivanov}, \citenamefont {Moore},
  \citenamefont {Singh}, \citenamefont {Rathnayake}, \citenamefont {Vig},
  \citenamefont {Granger}, \citenamefont {Muller}, \citenamefont {Bonazzi},
  \citenamefont {Gupta}, \citenamefont {Vats}, \citenamefont {Johansson},
  \citenamefont {Pedregosa}, \citenamefont {Curry}, \citenamefont {Terrel},
  \citenamefont {Rou\v{c}ka}, \citenamefont {Saboo}, \citenamefont {Fernando},
  \citenamefont {Kulal}, \citenamefont {Cimrman},\ and\ \citenamefont
  {Scopatz}}]{10.7717/peerj-cs.103}%
  \BibitemOpen
  \bibfield  {author} {\bibinfo {author} {\bibfnamefont {A.}~\bibnamefont
  {Meurer}}, \bibinfo {author} {\bibfnamefont {C.~P.}\ \bibnamefont {Smith}},
  \bibinfo {author} {\bibfnamefont {M.}~\bibnamefont {Paprocki}}, \bibinfo
  {author} {\bibfnamefont {O.}~\bibnamefont {\v{C}ert\'{i}k}}, \bibinfo
  {author} {\bibfnamefont {S.~B.}\ \bibnamefont {Kirpichev}}, \bibinfo {author}
  {\bibfnamefont {M.}~\bibnamefont {Rocklin}}, \bibinfo {author} {\bibfnamefont
  {A.}~\bibnamefont {Kumar}}, \bibinfo {author} {\bibfnamefont
  {S.}~\bibnamefont {Ivanov}}, \bibinfo {author} {\bibfnamefont {J.~K.}\
  \bibnamefont {Moore}}, \bibinfo {author} {\bibfnamefont {S.}~\bibnamefont
  {Singh}}, \bibinfo {author} {\bibfnamefont {T.}~\bibnamefont {Rathnayake}},
  \bibinfo {author} {\bibfnamefont {S.}~\bibnamefont {Vig}}, \bibinfo {author}
  {\bibfnamefont {B.~E.}\ \bibnamefont {Granger}}, \bibinfo {author}
  {\bibfnamefont {R.~P.}\ \bibnamefont {Muller}}, \bibinfo {author}
  {\bibfnamefont {F.}~\bibnamefont {Bonazzi}}, \bibinfo {author} {\bibfnamefont
  {H.}~\bibnamefont {Gupta}}, \bibinfo {author} {\bibfnamefont
  {S.}~\bibnamefont {Vats}}, \bibinfo {author} {\bibfnamefont {F.}~\bibnamefont
  {Johansson}}, \bibinfo {author} {\bibfnamefont {F.}~\bibnamefont
  {Pedregosa}}, \bibinfo {author} {\bibfnamefont {M.~J.}\ \bibnamefont
  {Curry}}, \bibinfo {author} {\bibfnamefont {A.~R.}\ \bibnamefont {Terrel}},
  \bibinfo {author} {\bibfnamefont {v.}~\bibnamefont {Rou\v{c}ka}}, \bibinfo
  {author} {\bibfnamefont {A.}~\bibnamefont {Saboo}}, \bibinfo {author}
  {\bibfnamefont {I.}~\bibnamefont {Fernando}}, \bibinfo {author}
  {\bibfnamefont {S.}~\bibnamefont {Kulal}}, \bibinfo {author} {\bibfnamefont
  {R.}~\bibnamefont {Cimrman}}, \ and\ \bibinfo {author} {\bibfnamefont
  {A.}~\bibnamefont {Scopatz}},\ }\href {\doibase 10.7717/peerj-cs.103}
  {\bibfield  {journal} {\bibinfo  {journal} {PeerJ Computer Science}\ }\textbf
  {\bibinfo {volume} {3}},\ \bibinfo {pages} {e103} (\bibinfo {year}
  {2017})}\BibitemShut {NoStop}%
\bibitem [{\citenamefont {Wright}\ \emph {et~al.}(1999)\citenamefont {Wright},
  \citenamefont {Nocedal} \emph {et~al.}}]{wright1999numerical}%
  \BibitemOpen
  \bibfield  {author} {\bibinfo {author} {\bibfnamefont {S.}~\bibnamefont
  {Wright}}, \bibinfo {author} {\bibfnamefont {J.}~\bibnamefont {Nocedal}},
  \emph {et~al.},\ }\href@noop {} {\bibfield  {journal} {\bibinfo  {journal}
  {Springer Science}\ }\textbf {\bibinfo {volume} {35}},\ \bibinfo {pages} {7}
  (\bibinfo {year} {1999})}\BibitemShut {NoStop}%
\bibitem [{\citenamefont {Mertens}(2000)}]{mertens2000random}%
  \BibitemOpen
  \bibfield  {author} {\bibinfo {author} {\bibfnamefont {S.}~\bibnamefont
  {Mertens}},\ }\href@noop {} {\bibfield  {journal} {\bibinfo  {journal}
  {Physical Review Letters}\ }\textbf {\bibinfo {volume} {84}},\ \bibinfo
  {pages} {1347} (\bibinfo {year} {2000})}\BibitemShut {NoStop}%
\bibitem [{\citenamefont {Smelyanskiy}\ \emph {et~al.}(2002)\citenamefont
  {Smelyanskiy}, \citenamefont {Toussaint},\ and\ \citenamefont
  {Timucin}}]{smelyanskiy2002dynamics}%
  \BibitemOpen
  \bibfield  {author} {\bibinfo {author} {\bibfnamefont {V.~N.}\ \bibnamefont
  {Smelyanskiy}}, \bibinfo {author} {\bibfnamefont {U.~v.}\ \bibnamefont
  {Toussaint}}, \ and\ \bibinfo {author} {\bibfnamefont {D.~A.}\ \bibnamefont
  {Timucin}},\ }\href@noop {} {\bibfield  {journal} {\bibinfo  {journal} {arXiv
  preprint quant-ph/0202155}\ } (\bibinfo {year} {2002})}\BibitemShut {NoStop}%
\bibitem [{\citenamefont {Barenco}\ \emph {et~al.}(1995)\citenamefont
  {Barenco}, \citenamefont {Bennett}, \citenamefont {Cleve}, \citenamefont
  {DiVincenzo}, \citenamefont {Margolus}, \citenamefont {Shor}, \citenamefont
  {Sleator}, \citenamefont {Smolin},\ and\ \citenamefont
  {Weinfurter}}]{barenco1995elementary}%
  \BibitemOpen
  \bibfield  {author} {\bibinfo {author} {\bibfnamefont {A.}~\bibnamefont
  {Barenco}}, \bibinfo {author} {\bibfnamefont {C.~H.}\ \bibnamefont
  {Bennett}}, \bibinfo {author} {\bibfnamefont {R.}~\bibnamefont {Cleve}},
  \bibinfo {author} {\bibfnamefont {D.~P.}\ \bibnamefont {DiVincenzo}},
  \bibinfo {author} {\bibfnamefont {N.}~\bibnamefont {Margolus}}, \bibinfo
  {author} {\bibfnamefont {P.}~\bibnamefont {Shor}}, \bibinfo {author}
  {\bibfnamefont {T.}~\bibnamefont {Sleator}}, \bibinfo {author} {\bibfnamefont
  {J.~A.}\ \bibnamefont {Smolin}}, \ and\ \bibinfo {author} {\bibfnamefont
  {H.}~\bibnamefont {Weinfurter}},\ }\href@noop {} {\bibfield  {journal}
  {\bibinfo  {journal} {Physical review A}\ }\textbf {\bibinfo {volume} {52}},\
  \bibinfo {pages} {3457} (\bibinfo {year} {1995})}\BibitemShut {NoStop}%
\bibitem [{\citenamefont {Saeedi}\ and\ \citenamefont
  {Pedram}(2013)}]{saeedi2013linear}%
  \BibitemOpen
  \bibfield  {author} {\bibinfo {author} {\bibfnamefont {M.}~\bibnamefont
  {Saeedi}}\ and\ \bibinfo {author} {\bibfnamefont {M.}~\bibnamefont
  {Pedram}},\ }\href@noop {} {\bibfield  {journal} {\bibinfo  {journal}
  {Physical Review A}\ }\textbf {\bibinfo {volume} {87}},\ \bibinfo {pages}
  {062318} (\bibinfo {year} {2013})}\BibitemShut {NoStop}%
\bibitem [{\citenamefont {He}\ \emph {et~al.}(2017)\citenamefont {He},
  \citenamefont {Luo}, \citenamefont {Zhang}, \citenamefont {Wang},\ and\
  \citenamefont {Wang}}]{he2017decompositions}%
  \BibitemOpen
  \bibfield  {author} {\bibinfo {author} {\bibfnamefont {Y.}~\bibnamefont
  {He}}, \bibinfo {author} {\bibfnamefont {M.-X.}\ \bibnamefont {Luo}},
  \bibinfo {author} {\bibfnamefont {E.}~\bibnamefont {Zhang}}, \bibinfo
  {author} {\bibfnamefont {H.-K.}\ \bibnamefont {Wang}}, \ and\ \bibinfo
  {author} {\bibfnamefont {X.-F.}\ \bibnamefont {Wang}},\ }\href@noop {}
  {\bibfield  {journal} {\bibinfo  {journal} {International Journal of
  Theoretical Physics}\ }\textbf {\bibinfo {volume} {56}},\ \bibinfo {pages}
  {2350} (\bibinfo {year} {2017})}\BibitemShut {NoStop}%
\bibitem [{\citenamefont {Wolpert}\ and\ \citenamefont
  {Macready}(1997)}]{wolpert1997no}%
  \BibitemOpen
  \bibfield  {author} {\bibinfo {author} {\bibfnamefont {D.~H.}\ \bibnamefont
  {Wolpert}}\ and\ \bibinfo {author} {\bibfnamefont {W.~G.}\ \bibnamefont
  {Macready}},\ }\href@noop {} {\bibfield  {journal} {\bibinfo  {journal} {IEEE
  transactions on evolutionary computation}\ }\textbf {\bibinfo {volume} {1}},\
  \bibinfo {pages} {67} (\bibinfo {year} {1997})}\BibitemShut {NoStop}%
\bibitem [{\citenamefont {Yan}\ and\ \citenamefont
  {Sinitsyn}(2022)}]{yan2022analytical}%
  \BibitemOpen
  \bibfield  {author} {\bibinfo {author} {\bibfnamefont {B.}~\bibnamefont
  {Yan}}\ and\ \bibinfo {author} {\bibfnamefont {N.~A.}\ \bibnamefont
  {Sinitsyn}},\ }\href@noop {} {\bibfield  {journal} {\bibinfo  {journal}
  {Nature Communications}\ }\textbf {\bibinfo {volume} {13}},\ \bibinfo {pages}
  {1} (\bibinfo {year} {2022})}\BibitemShut {NoStop}%
\bibitem [{\citenamefont {Morley}\ \emph {et~al.}(2019)\citenamefont {Morley},
  \citenamefont {Chancellor}, \citenamefont {Bose},\ and\ \citenamefont
  {Kendon}}]{morley2019quantum}%
  \BibitemOpen
  \bibfield  {author} {\bibinfo {author} {\bibfnamefont {J.~G.}\ \bibnamefont
  {Morley}}, \bibinfo {author} {\bibfnamefont {N.}~\bibnamefont {Chancellor}},
  \bibinfo {author} {\bibfnamefont {S.}~\bibnamefont {Bose}}, \ and\ \bibinfo
  {author} {\bibfnamefont {V.}~\bibnamefont {Kendon}},\ }\href@noop {}
  {\bibfield  {journal} {\bibinfo  {journal} {Physical review A}\ }\textbf
  {\bibinfo {volume} {99}},\ \bibinfo {pages} {022339} (\bibinfo {year}
  {2019})}\BibitemShut {NoStop}%
\bibitem [{\citenamefont {Jiang}\ \emph
  {et~al.}(2017{\natexlab{b}})\citenamefont {Jiang}, \citenamefont {Rieffel},\
  and\ \citenamefont {Wang}}]{jiang2017near}%
  \BibitemOpen
  \bibfield  {author} {\bibinfo {author} {\bibfnamefont {Z.}~\bibnamefont
  {Jiang}}, \bibinfo {author} {\bibfnamefont {E.~G.}\ \bibnamefont {Rieffel}},
  \ and\ \bibinfo {author} {\bibfnamefont {Z.}~\bibnamefont {Wang}},\
  }\href@noop {} {\bibfield  {journal} {\bibinfo  {journal} {Physical Review
  A}\ }\textbf {\bibinfo {volume} {95}},\ \bibinfo {pages} {062317} (\bibinfo
  {year} {2017}{\natexlab{b}})}\BibitemShut {NoStop}%
\bibitem [{\citenamefont {Roland}\ and\ \citenamefont
  {Cerf}(2002)}]{roland2002quantum}%
  \BibitemOpen
  \bibfield  {author} {\bibinfo {author} {\bibfnamefont {J.}~\bibnamefont
  {Roland}}\ and\ \bibinfo {author} {\bibfnamefont {N.~J.}\ \bibnamefont
  {Cerf}},\ }\href@noop {} {\bibfield  {journal} {\bibinfo  {journal} {Physical
  Review A}\ }\textbf {\bibinfo {volume} {65}},\ \bibinfo {pages} {042308}
  (\bibinfo {year} {2002})}\BibitemShut {NoStop}%
\bibitem [{\citenamefont {Callison}\ \emph {et~al.}(2019)\citenamefont
  {Callison}, \citenamefont {Chancellor}, \citenamefont {Mintert},\ and\
  \citenamefont {Kendon}}]{callison2019finding}%
  \BibitemOpen
  \bibfield  {author} {\bibinfo {author} {\bibfnamefont {A.}~\bibnamefont
  {Callison}}, \bibinfo {author} {\bibfnamefont {N.}~\bibnamefont
  {Chancellor}}, \bibinfo {author} {\bibfnamefont {F.}~\bibnamefont {Mintert}},
  \ and\ \bibinfo {author} {\bibfnamefont {V.}~\bibnamefont {Kendon}},\
  }\href@noop {} {\bibfield  {journal} {\bibinfo  {journal} {New Journal of
  Physics}\ }\textbf {\bibinfo {volume} {21}},\ \bibinfo {pages} {123022}
  (\bibinfo {year} {2019})}\BibitemShut {NoStop}%
\end{thebibliography}%
\bibliographystyle{apsrev4-1}
\clearpage
\newpage
\widetext    
\begin{appendix}
\section{Permutation Symmetry of Problem Hamiltonian Under Grover-QAOA State}
\label{permutation_independence_of_expectation}
The invariance of a QAOA problem Hamiltonian under permutation operator \begin{equation}    
    \hat P = \hat U_{j_m \leftrightarrow k_m}...\hat U_{j_2 \leftrightarrow k_2}\hat U_{j_1 \leftrightarrow k_1}
    \end{equation} 
can be demonstrated using the identities
\begin{equation}
    \hat P \hat P^{\dagger} = \hat I, \,\, e^{\hat P^{\dagger}\hat O \hat P} = \hat P^{\dagger}e^{\hat O }\hat P,
\end{equation}
therefore:
    \begin{multline}
        \langle \hat P^\dagger \hat H_{\textrm{P}} \hat P \rangle _{\vec \beta,\vec\gamma}= \bra +^{\otimes n}\prod_{p'=1}^p  \left[e^{i\gamma_{p'} \hat P^\dagger \hat H_{\rm P} \hat P}\hat e^{i\beta_{p'} \hat H_{\rm D}}\right]P^\dagger \hat H_{\rm P} \hat P\prod_{p'=1}^p \left[ e^{i\beta_{p'}\hat H_{\rm D}} e^{i\gamma_{p'} \hat P^\dagger \hat H_{\rm P} \hat P}\right]\ket+^{\otimes n}\\
        = \bra +^{\otimes n}\prod_{p'=1}^p  \left[\hat P^\dagger e^{i\gamma_{p'}  \hat H_{\rm P} }\hat P\hat e^{i\beta_{p'} \hat H_{\rm D}}\right]P^\dagger \hat H_{\rm P} \hat P\prod_{p'=1}^p \left[ e^{i\beta_{p'}\hat H_{\rm D}}\hat P^\dagger e^{i\gamma_{p'}  \hat H_{\rm P} }\hat P\right]\ket+^{\otimes n}
        \\
        = \bra +^{\otimes n}P^\dagger\prod_{p'=1}^p  \left[\hat  e^{i\gamma_{p'}  \hat H_{\rm P} }\hat Pe^{i\beta_{p'} \hat H_{\rm D}}\hat  P^\dagger \right] \hat H_{\rm P} \prod_{p'=1}^p \left[\hat P e^{i\beta_{p'}\hat H_{\rm D}}\hat P^\dagger e^{i\gamma_{p'}  \hat H_{\rm P} }\right]\hat P\ket+^{\otimes n}
        \\
        = \bra +^{\otimes n}P^\dagger\prod_{p'=1}^p  \left[\hat  e^{i\gamma_{p'}  \hat H_{\rm P} } e^{i\beta_{p'}\hat P \hat H_{\rm D}\hat P^\dagger}  \right] \hat H_{\rm P} \prod_{p'=1}^p \left[ e^{i\beta_{p'}\hat P\hat H_{\rm D}\hat P^\dagger} e^{i\gamma_{p'}  \hat H_{\rm P} }\right]\hat P\ket+^{\otimes n}
        \\
        =\langle \hat H_{\textrm{P}} \rangle _{\vec \beta,\vec\gamma}
    \end{multline}
    demonstrating that the expectation value is invariant under any permutation.
    
\section{Calculation of Depth $1$ Expectation Value}
\label{depth1_appendix_section}

We can write down the unitaries for the problem Hamiltonian:
\begin{equation}
\hat U_P(\gamma) = \sum_z \mathrm{e}^{i\gamma C(z)} \ket z\bra z
\end{equation}
and driver Hamiltonian:
\begin{equation}
\hat U_D(\beta) =  \mathrm{e}^{i\beta \ket{+}\bra{+}} = \mathcal I + i\beta \ket{+}\bra{+} +\frac{(i\beta)^2}{2}\ket{+}\bra{+} ...
=  \mathcal I + (\mathrm{e}^{i\beta} - 1)\ket{+}\bra{+} =  \mathcal I + B(\beta)\ket{+}\bra{+}
\end{equation}
with $B(\beta) = e^{i\beta} - 1$. For $p = 1$ the problem expectation can be written as:
\begin{multline}
\langle \hat H_P \rangle_{\gamma, \beta} = \bra{+} \hat U_\mathrm{P}^\dagger(\gamma)\hat U_\mathrm{D}^\dagger(\beta)\hat H_P \hat U_\mathrm{D}(\beta)\hat U_\mathrm{P}(\gamma) \ket{+} \\
= \sum_{z_{-1},z_{0},z_{1}}\bra{+}  \mathrm{e}^{-i\gamma C(z_{-1})} \ket{z_{-1}}\bra{z_{-1}} \big[\mathcal{I} + B^*\ket{+}\bra{+}\big]
C(z_{0})\ket{z_{0}}\bra{z_{0}} \big[\mathcal I + B\ket{+}\bra{+}\big]\mathrm{e}^{i\gamma C(z_{1})} \ket{z_{1}}\bra{z_{1}}) \ket{+} \\
= \frac1{N}  \sum_{z_{-1},z_{0},z_{1}} \mathrm{e}^{-i\gamma C(z_{-1})} \bra{z_{-1}} \big[\mathcal I + B^*\ket{+}\bra{+}\big] C(z_{0})\ket{z_{0}}\bra{z_{0}} \big[\mathcal I + B\ket{+}\bra{+}\big]\mathrm{e}^{i\gamma C(z_{1})}\ket{z_{1}}
\end{multline}
Which yields four terms as follows:
\begin{multline}
\bar C +\frac{1}{N^2}BB^*\bar C \sum_{z} \mathrm{e}^{i\gamma C(z)}\sum_{z}\mathrm{e}^{-i\gamma C(z)}
+\frac1{N^2}B \sum_{z}\mathrm{e}^{i\gamma C(z)}
\sum_{z}C(z)\mathrm{e}^{-i\gamma C(z)} +\frac1{N^2}B^*\sum_{z}
\mathrm{e}^{-i\gamma C(z)} \sum_{z} C(z) \mathrm{e}^{i\gamma C(z)}.
\end{multline}
One then can substitute the characteristic function of the ensemble to obtain an expression for the expectation value of the QAOA state, with 
\begin{equation}
        \frac{1}{N}\sum_{z} C(z) \to \bar C, \,\,
 \sum_z \frac{e^{i\gamma C(z)}}{N}\to \Gamma(\gamma) 
\end{equation}
To express the term containing $C(z)$ not in the exponent, one may differentiate under the sum/integral as        
\begin{equation}
    \sum_{z}C(z)\mathrm{e}^{i\gamma C(z)} = -i\frac{\mathrm{d}}{\mathrm{d}\gamma} \sum_{z}C(z)\mathrm{e}^{-i\gamma C(z)}
    \to -i\frac{\mathrm{d}}{\mathrm{d}\gamma'} \Gamma(\gamma') \bigg\vert_{\gamma} = -i\Gamma'(\gamma).
\end{equation}
Resulting in the expression: 
\begin{equation}
\langle \hat H_P \rangle_{\gamma, \beta} = \bar C(1 + BB^*\Gamma \Gamma^*) + 2\Im(B^*\Gamma^* \Gamma') .
\end{equation}

\section{Calculation of Depth $2$ Expectation Value}

The $p =2$ expression can calculated similarly to the $p=1$ expression, however, with more terms and thus requiring the use of a computer algebra package \cite{10.7717/peerj-cs.103}. The result of expanding the expression for the expectation value, setting the mean $\bar C$ to zero, and substituting the characteristic function and its derivative is the following expression of 10 terms:
\begin{multline}
    E_2(\vec\gamma,\vec\beta) = i B{\left(\beta_{1} \right)} B{\left(\beta_{2} \right)} \Gamma{\left(\gamma_{1} \right)} \Gamma{\left(\gamma_{2} \right)} B{\left(\beta_{1} \right)}^{*} \Gamma{\left(\gamma_{1} \right)}^{*}  \Gamma ' {\left(\gamma_{2} \right)}^{*} 
     - i B{\left(\beta_{1} \right)} \Gamma{\left(\gamma_{1} \right)} B{\left(\beta_{1} \right)}^{*} B{\left(\beta_{2} \right)}^{*} \Gamma{\left(\gamma_{1} \right)}^{*} \Gamma{\left(\gamma_{2} \right)}^{*}  \Gamma ' {\left(\gamma_{2} \right)} \\
    - i B{\left(\beta_{1} \right)}^{*} \Gamma{\left(\gamma_{1} \right)}^{*}  \Gamma ' {\left(\gamma_{1} \right)}
    + i B{\left(\beta_{1} \right)} \Gamma{\left(\gamma_{1} \right)}  \Gamma ' {\left(\gamma_{1} \right)}^{*} 
    + i B{\left(\beta_{1} \right)} B{\left(\beta_{2} \right)} \Gamma{\left(\gamma_{1} \right)} \Gamma{\left(\gamma_{2} \right)} \Gamma '(\gamma_{1} + \gamma_{2})^{*} - i B{\left(\beta_{1} \right)}^{*} B{\left(\beta_{2} \right)}^{*} \Gamma{\left(\gamma_{1} \right)}^{*} \Gamma{\left(\gamma_{2} \right)}^{*}  \Gamma '(\gamma_{1} + \gamma_{2})\\
    - i B{\left(\beta_{1} \right)} \Gamma{\left(\gamma_{1} \right)} B{\left(\beta_{2} \right)}^{*}   \Gamma ' {\left(\gamma_{2} \right)}\Gamma{\left(\gamma_{1} + \gamma_{2} \right)}^{*} + i B{\left(\beta_{2} \right)}  B{\left(\beta_{1} \right)}^{*} \Gamma{\left(\gamma_{1} \right)}^{*}  \Gamma ' {\left(\gamma_{2} \right)}^{*} \Gamma{\left(\gamma_{1} + \gamma_{2} \right)}\\
    - i B{\left(\beta_{2} \right)}^{*} \Gamma{\left(\gamma_{1} + \gamma_{2} \right)}^{*}  \Gamma '(\gamma_{1} + \gamma_{2})
    + i B{\left(\beta_{2} \right)} \Gamma{\left(\gamma_{1} + \gamma_{2} \right)} \Gamma '(\gamma_{1} + \gamma_{2})^{*}.
\end{multline}
This expression can be further simplified via the substitution of the $p =1$ expectation resulting in the final expression in equation \ref{depth2_final_expression}.

\section{Calculation of Arbitrary Depth Expectation Value}
\label{arbitrary_depth_calculation}

Using the intuition from the $p = 1$ and $p = 2$ cases, we can write down the expression for any term in the arbitrary depth expectation value of the problem Hamiltonian. We can write the full expectation $E_p(\vec \gamma, \vec \beta)$ as:
\begin{equation}
    \bra{\vec \gamma, \vec \beta} \hat H_P\ket{\vec \gamma, \vec \beta} =     \bra{+}\prod_{j=-p} ^ {-1} \hat U_\mathrm{P}(\gamma_j)\hat U_\mathrm{D}(\beta_j) \hat H_{\rm P} \prod_{i=1} ^ {p}\hat U_\mathrm{D}(\beta_i)\hat U_\mathrm{P}(\gamma_i) \ket{+}    
\end{equation}
where we have introduced the convention that negative indices simply add a negative sign to the value, as: \begin{equation}
    \gamma_{-i} = -\gamma_i, \,\, \beta_{-i} = -\beta_i.
\end{equation} The expectation value can be expanded, substituting the expressions for the driver and problem Hamiltonians as:
\begin{equation}
\sum_{\vec z} \bra{+}\prod_{j= -1} ^ {-p} \Big[ e^{i\gamma_j C(z_j)}\ket{z_j}\bra{z_j}  \big(I+B_j\ket+\bra+\big)\Big] C(z_0)\ket{z_0}\bra{z_0} \prod_{i=1} ^ p \Big[\big(I+B_i\ket+\bra+\big) e^{i\gamma_i C(z_i)}\ket{z_i}\bra{z_i}  \Big] \ket{+}.  
\end{equation}
Where the sum over $2p + 1$ indices $\vec{z} = (z_{-p},...,z_{0},...,z_{p})$ runs over the set $\{0,N-1\}$ for each index. Expanding the products in this expression will yield $2^{2p}$ terms that can be sorted in the number of occurrences of a $B$ variable in the bra or ket. 

For the one term with zero $B$ incidences, one finds, for example:
\begin{equation}
\frac{1}{N}\sum_{\vec z} e^{\sum_{j= -1} ^ {-p} i\gamma_j C(z_j)}C(z_0)e^{\sum_i ^ p i\gamma_i C(z_i)}\bra{z_{-1}}\ket{z_{-2}}\bra{z_{-2}}...\ket{z_{-p}}\bra{z_{-p}}  \ket{z_0}\bra{z_0} \ket{z_p}\bra{z_p}...\ket{z_{2}}\bra{z_{2}}\ket{z_1}
\end{equation}
in which the orthogonal bra-kets evaluate to zero leaving only one index yielding a non-zero term evaluating to the mean of the objective function $\bar C$. For weight-$1$ terms one has $B$ in a position $l$
\begin{equation}
\sum_{\vec z} \bra{+}\prod_{j= -1} ^ {-p} \Big[ e^{-i\gamma C(z_j)}\ket{z_j}\bra{z_j}  \Big]C(z)\ket{z_0}\bra{z_0}  \prod_{i=l} ^ p \Big[ \ket+\bra+ e^{i\gamma C(z_i)}\ket{z_i}\bra{z_i} \Big]B_l\prod_{i=1} ^ {l-1} \Big[ \ket+\bra+ e^{i\gamma C(z_i)}\ket{z_i}\bra{z_i} \Big] \ket{+}    
\end{equation}
for which the orthogonal bra-kets can be evaluated as:
\begin{equation}
\frac{1}{N^2}\sum_{\vec z} e^{\sum_{j= -1} ^ {-p}i\gamma_j C(z_j)}C(z_0)e^{\sum_i ^ p -i\gamma_i C(z_i)} B_k
\delta_{z_{-1},z_{-2}}...\delta_{z_{-p},z_{0}}\delta_{z_{0},z_{p}}...\delta_{z_{k+2},z_{k+1}}\delta_{z_{k},z_{k-1}}...\delta_{z_{2},z_{1}} 
\end{equation}
in which the sum is eliminated for all but two partitions giving:
\begin{equation}
\frac{1}{N^2}\sum_{z}e^{\sum_{j= -1} ^ {-p}i\gamma_j C(z)}C(z)e^{\sum_{i=k+1} ^ p i\gamma_i C(z)}\sum_{z} e^{\sum_{i= 1} ^ k i\gamma_i C(z)} B_k
\end{equation}
leaving two variables over which we sum and leaving the non-exponentiated $C(z)$ and therefore the differentiated characteristic function $\Gamma'$ in the factor corresponding to the partition containing the central variable $z_0$
\begin{equation}
-iB_k \Gamma'\left(\sum_{i= -1} ^ {-p}\gamma_i + \sum_{i=k+1} ^ p \gamma_i\right) \Gamma\left(\sum_{i= 1} ^ k \gamma_i\right).
\end{equation}

To expand the expression more generally, we can introduce indices $k_\mathrm{bra} \in \{0...2^p-1\}$ and $k_\mathrm{ket}\in \{0...2^p-1\}$ that are each $p$-bit numbers such that the presence of a $1$ in the binary representation of the index implies a factor of $B_i\ket+\bra+$ at the bit position in the bra or ket for the relevant term. The formula can be rewritten as
\begin{equation}
\sum_{k_{\mathrm{bra}},\,k_\mathrm{ket} = 0}^{2^p-1}\sum_{\vec z} \bra{+}\prod_{j= -p} ^ {-1} \Big[ e^{i\gamma_j C(z_j)}\ket{z_j}\bra{z_j}  \big(B_j\ket+\bra+\big)^{k_{\mathrm{bra}}^j}\Big] C(z_0)\ket{z_0}\bra{z_0} \prod_{i=1} ^ p \Big[\big(B_i\ket+\bra+\big)^{k_{\mathrm{ket}}^i} e^{i\gamma_i C(z_i)}\ket{z_i}\bra{z_i}  \Big] \ket{+}.  
\end{equation}
When the index $k_{ket/bra}$ is zero, this expression $B_i\ket+\bra+$ is the identity and adjacent non-identical $z$ indices result in a Kronecker delta, with $\bra{z_j}\ket{z_k} = \delta_{z_j,z_k}$ collapsing a sum over one of the $z$ indices. As such, these terms are collected into a single sum. When a bit in the index $k_{ket/bra}$ is one, the projector $\ket+\bra+$ annihilates with a basis state vector $\bra{z_i}$ and simply produces a factor of $1/N$, with $\bra+\ket{z} = 1/\sqrt{N}\,\, \forall z$, leaving the sums over $z$ intact. 

To substitute the characteristic function, we note that unbroken chains of $0$'s of length $n^*$ in the indices $k_{ket/bra}$ will result in factors of the form:
\begin{multline}
    \sum_{z_1, z_2... z_{n^*}} e^{i\gamma_1 C(z_1)}\ket{z_1}\bra{z_1}e^{i\gamma_2 C(z_2)}\ket{z_2}\bra{z_2}...e^{i\gamma_{n^*} C(z_{n^*})}\ket{z_{n^*}}\bra{z_{n^*}}\\ =\sum_{z_1, z_2... z_{n^*}} \delta_{z_1,z_2}\delta_{z_2,z_3}...\delta_{z_{n^*-1},z_{n^*}}e^{i\gamma_1 C(z_1)}e^{i\gamma_2 C(z_2)}...e^{i\gamma_{n^*} C(z_{n^*})}\\
    = \sum_{z} e^{iC(z)\sum_{i\in{[1...n^*]}}\gamma_i }
    = \Gamma\left(\sum_{i\in{[1...n^*]}}\gamma_i\right).
\end{multline}
To substitute the characteristic function for all incidences of the objective function $C(z)$, it is convenient to define a set of boundary indices between those that are grouped as above. We can define these as 
\begin{equation}
S_\mathrm{bra} =  \{0\} + \{i \vert  k_\mathrm{bra}^i = 1\}, \,\, S_\mathrm{ket} = \{0\} + \{i \vert  k_\mathrm{ket}^i = 1\}
\end{equation} defined by the locations of $1$'s in the binary representation of $k_\mathrm{bra}$, $k_\mathrm{ket}$. The additional boundary at $0$ represents the boundary imposed by the initial state of QAOA. In the partitions formed by these boundaries, we collect the indices contained in the $i$th partition of the bra/ket as $P_\mathrm{bra}^{i}$, $P_\mathrm{ket}^{i}$ with 
\begin{equation}
P_\mathrm{bra}^{i} = \{-j \vert j > k_\mathrm{bra,}^i, j \leq k_\mathrm{bra,}^{i+1}\}, P_\mathrm{ket}^{i} = \{j \vert j > k_\mathrm{ket,}^i, j \leq k_\mathrm{ket,}^{i+1}\}.
\end{equation}
And collect these into a set of sets
\begin{equation}
P_\mathrm{bra} = \{P_\mathrm{bra}^i \vert i \in [0,\mathrm{Weight}(k_{\mathrm{bra}})]\}, P_\mathrm{ket} = \{P_\mathrm{ket}^i \vert i \in [0,\mathrm{Weight}(k_{\mathrm{ket}})]\}.
\end{equation}
Finally, one defines a central partition spanning the indices between the last occurence $1$ in the binary representation of $k_\mathrm{bra}$, $k_\mathrm{ket}$ and containing the non-exponentiated factor of the objective function 
\begin{equation}
    P_{\mathrm{central}} = \{-j \vert j > \max S_\mathrm{bra}\} + \{j \vert j > \max S_\mathrm{ket}\}.
\end{equation} 
The expectation value after the substitution of the characteristic function and its derivative can then be expressed as:
\begin{equation}
        E_p(\vec \gamma, \vec \beta) = \sum_{k_\mathrm{bra}, k_\mathrm{ket}=0}^{2^p-1}-i \prod_{P \in P_{\mathrm{bra}}} \Gamma\left(\sum_{i\in P} \gamma_i \right)\Gamma'\left(\sum_{i\in P_{\rm central}} \gamma_i \right)\prod_{P \in P_{\mathrm{ket}}} \Gamma\left(\sum_{i\in P} \gamma_i \right)\prod_{-j \vert k_{\mathrm{bra}}^j=1} B_j\prod_{j \vert k_{\mathrm{ket}}^j=1} B_j
\end{equation}
For which exchanging $k_\mathrm{bra}$, $k_\mathrm{ket}$ conjugates the term, allowing the expression to be split into two parts, one in which $k_\mathrm{bra} = k_\mathrm{ket}$ and one otherwise, so:
\begin{multline}
        E_p(\vec \gamma, \vec \beta) = \sum_{k=0}^{2^p-1}-i \prod_{P \in P_{\mathrm{bra}}} \Gamma\left(\sum_{i\in P} \gamma_i \right)\Gamma\left(\sum_{i\in P} \gamma_i \right)^*\Gamma'\left(0\right)\prod_{j \vert k^j=1} B_jB_j^*\\
        +  \sum_{k_\mathrm{bra}< k_\mathrm{ket}=0}^{2^p-1}-i \prod_{P \in P_{\mathrm{bra}}} \Gamma\left(\sum_{i\in P} \gamma_i \right)\Gamma'\left(\sum_{i\in P_{\rm central}} \gamma_i \right)\prod_{P \in P_{\mathrm{ket}}} \Gamma\left(\sum_{i\in P} \gamma_i \right)\prod_{-j \vert k_{\mathrm{bra}}^j=1} B_j\prod_{j \vert k_{\mathrm{ket}}^j=1} B_j  + \mathrm{Conjugate}.
\end{multline}
If the problem mean is zeroed by global phase, $\Gamma'(0) = 0$ and the first sum disappears, leaving:
\begin{equation}
        E_p(\vec \gamma, \vec \beta) =  2\Im{\sum_{k_\mathrm{bra}< k_\mathrm{ket}=0}^{2^p-1}\prod_{P \in P_{\mathrm{bra}}} \Gamma\left(\sum_{i\in P} \gamma_i \right)\Gamma'\left(\sum_{i\in P_{\rm central}} \gamma_i \right)\prod_{P \in P_{\mathrm{ket}}} \Gamma\left(\sum_{i\in P} \gamma_i \right)\prod_{-j \vert k_{\mathrm{bra}}^j=1 \, \mathrm{or}\, j \vert k_{\mathrm{ket}}^j=1} B_j}.
\end{equation}
A python script to evaluate the Grover-QAOA objective function $E_p(\vec \gamma, \vec \beta)$ for a given characteristic function is available upon request to the author.

\end{appendix}

\end{document}